\documentclass[twocolumn]{aastex631}
\usepackage{sidecap}
\usepackage{wrapfig}
\usepackage{float}
\usepackage{xcolor}
\usepackage{hyperref}

\usepackage{tablefootnote}

\shorttitle{WASP-18\,$\mathrm{b}$ Dayside}
\shortauthors{Brogi et al.}
\graphicspath{{./}}

\begin{document}
\title{The Roasting Marshmallows Program with IGRINS on Gemini South I: Composition and Climate of the Ultra Hot Jupiter WASP-18\,b}

\correspondingauthor{Matteo Brogi}
\email{m.brogi@warwick.ac.uk}

\author[0000-0002-7704-0153]{Matteo Brogi}
\affil{Department of Physics, University of Warwick, Coventry CV4 7AL, UK}
\affil{INAF-Osservatorio Astrofisico di Torino, Via Osservatorio 20,
I-10025 Pino Torinese, Italy}
\affil{Centre for Exoplanets and Habitability, University of Warwick, Coventry, CV4 7AL, UK}

\author{Vanessa Emeka-Okafor}
\affil{Department of Physics, University of Warwick, Coventry CV4 7AL, UK}

\author[0000-0001-6247-8323]{Michael R.\ Line}
\affil{School of Earth and Space Exploration, Arizona State University, Tempe, AZ 85281, USA}

\author{Siddharth Gandhi}
\affil{Leiden Observatory, Leiden University, Postbus 9513, 2300 RA Leiden, The Netherlands}
\affil{Department of Physics, University of Warwick, Coventry CV4 7AL, UK}
\affil{Centre for Exoplanets and Habitability, University of Warwick, Coventry, CV4 7AL, UK}

\author{Lorenzo Pino}
\affil{INAF-Osservatorio Astrofisico di Arcetri, Largo Enrico Fermi 5, 50125 Firenze}

\author{Eliza M.-R. Kempton}
\affil{Department of Astronomy, University of Maryland, 4296 Stadium Drive, College Park, MD 20742, USA}

\author{Emily Rauscher}
\affil{Department of Astronomy, University of Michigan, Ann Arbor, MI 48109, USA}

\author{Vivien Parmentier}
\affil{Universit\'e C\^ote d'Azur, Observatoire de la C\^ote d'Azur, CNRS, Laboratoire Lagrange, France}
\affil{Atmospheric, Oceanic, and Planetary Physics, Clarendon Laboratory, Department of Physics, University of Oxford, Oxford OX1 3PU, UK}

\author{Jacob L. Bean}
\affil{Department of Astronomy and Astrophysics, University of Chicago, 5640 South Ellis Avenue,  Chicago, IL 60637, USA }

\author{Gregory N. Mace}
\affil{Department of Astronomy, University of Texas at Austin, 2515 Speedway, Austin, TX, USA }

\author[0000-0001-6129-5699]{Nicolas B. Cowan}
\affil{Department of Earth \& Planetary Sciences, McGill University, 3450 rue University, Montr\'eal, QC H3A 0E8, Canada}
\affil{Department of Physics, McGill University, 3600 rue University, Montr\'eal, QC H3A 2T8, Canada}

\author{Evgenya Shkolnik}
\affil{School of Earth and Space Exploration, Arizona State University, Tempe, AZ 85281, USA}

\author{Joost P. Wardenier}
\affil{Atmospheric, Oceanic, and Planetary Physics, Clarendon Laboratory, Department of Physics, University of Oxford, Oxford OX1 3PU, UK}

\author{Megan Mansfield}
\affil{Steward Observatory, University of Arizona, Tucson, AZ 85715, USA}
\affil{NHFP Sagan Fellow}

\author{Luis Welbanks}
\affil{School of Earth and Space Exploration, Arizona State University, Tempe, AZ 85281, USA}
\affil{NHFP Sagan Fellow}

\author{Peter Smith}
\affil{School of Earth and Space Exploration, Arizona State University, Tempe, AZ 85281, USA}

\author{Jonathan J. Fortney}
\affil{Department of Astronomy and Astrophysics, University of California, Santa Cruz, CA 95064,USA}

\author{Jayne L. Birkby}
\affil{Astrophysics, University of Oxford, Denys Wilkinson Building, Keble Road, Oxford OX1 3RH, UK}

\author[0000-0002-2259-4116]{Joseph A. Zalesky}
\affil{Department of Astronomy, University of Texas at Austin, 2515 Speedway, Austin, TX 78712, USA}

\author{Lisa Dang}
\affil{Department of Physics, McGill University, 3600 rue University, Montr\`eal, QC H3A 2T8, Canada}

\author{Jennifer Patience}
\affil{School of Earth and Space Exploration, Arizona State University, Tempe, AZ 85281, USA}

\author{Jean-Michel D{\'{e}}sert}
\affil{Anton Pannekoek Institute for Astronomy, University of Amsterdam, Science Park 904, 1098 XH Amsterdam, The Netherlands}

\begin{abstract}
We present high-resolution dayside thermal emission observations of the exoplanet WASP-18\,b using IGRINS on Gemini South. We remove stellar and telluric signatures using standard algorithms, and we extract the planet signal via cross correlation with model spectra. We detect the atmosphere of WASP-18\,b at a signal-to-noise ratio (SNR) of 5.9 using a full chemistry model, measure H$_2$O (SNR=3.3), CO (SNR=4.0), and OH (SNR=4.8) individually, and confirm previous claims of a thermal inversion layer. The three species are confidently detected ($>$4$\sigma$) with a Bayesian inference framework, which we also use to retrieve abundance, temperature, and velocity information. For this ultra-hot Jupiter (UHJ), thermal dissociation processes likely play an important role. Retrieving abundances constant with altitude and allowing the temperature-pressure profile to freely adjust results in a moderately super-stellar carbon to oxygen ratio (C/O=$0.75^{+0.14}_{-0.17}$) and metallicity ([M/H]=$1.03^{+0.65}_{-1.01}$). Accounting for undetectable oxygen produced by thermal dissociation leads to C/O=$0.45^{+0.08}_{-0.10}$ and [M/H]$=1.17^{+0.66}_{-1.01}$. A retrieval that assumes radiative-convective-thermochemical-equilibrium and naturally accounts for thermal dissociation constrains C/O$<$0.34 (2$\sigma$) and [M/H]$=0.48^{+0.33}_{-0.29}$, in line with the chemistry of the parent star. Looking at the velocity information, we see a tantalising signature of different Doppler shifts at the level of a few km s$^{-1}$ for different molecules, which might probe dynamics as a function of altitude and/or location on the planet disk. Our results demonstrate that ground-based, high-resolution spectroscopy at infrared wavelengths can provide meaningful constraints on the compositions and climate of highly irradiated planets. This work also elucidates potential pitfalls with commonly employed retrieval assumptions when applied to the spectra of UHJs.  
\end{abstract}


\section{Introduction} \label{sec:intro}
Atmospheric characterization of the exoplanet population provides insight into the mechanisms that drive planetary climate, atmospheric chemical processes and transitions, and planet formation processes \citep{Madhusudhan19, ZhangXi2020}.   Transit spectroscopy provides an avenue for measuring diagnostic quantities, such as vertical/horizontal temperature structure and atmospheric composition (volume mixing ratios of individual molecules and atoms/ions) that enable the understanding these critical mechanisms, processes, and transitions \citep{SeagerBook2010, Madhu2018, Guillot2022}, especially if such measurements exist over a large sample of planets \citep{Baxter2020,Welbanks2019, Mansfield2021}.

Most of the spectroscopic observations either during transit or secondary eclipse use space based platforms like HST and Spitzer (and currently JWST) with low-to-moderate spectral resolution (R$\sim$50-1000) or photometry-based instruments \citep[primarily HST WFC3/STIS and Spitzer IRAC,][]{Sing2016, Guillot2022}. A complementary approach to space-based transit spectroscopy is ground-based time-series with high resolution cross-correlation spectroscopy \citep[HRCCS,][]{Snellen2010,Birkby2018}. HRCCS utilizes the planetary orbital Doppler shift of large numbers of molecular lines  attainable at high spectral resolutions ($R$\,$\gtrapprox$\,15,000) to disentangle the planetary atmosphere signal from telluric and stellar contaminants. The high spectral resolution is particularly sensitive to the presence of molecular and atomic line absorbers as well as to temperature gradients \citep{BrogiBirkby2021}. Leveraging this sensitivity, \cite{BL19} showed that quantitative constraints on the key diagnostic quantities--the vertical temperature structure (temperature vs. pressure, TP) and gas volume mixing ratios--could be retrieved directly from these types of data (e.g., with the CRIRES instrument on the VLT) with precision potentially exceeding that achieved when applying similar methods to HST and Spitzer transit observations.    

Motivated by the potential for HRCCS observations  with modern instruments to provide ultra-precise atmospheric temperature and absolute abundance constraints,  \cite{line2021} leveraged the broad, simultaneous wavelength coverage (1.43-2.42 $\mu$m) and high spectral resolution (R$\approx$45,000) of IGRINS \citep[Immersion GRating INfrared Spectrometer,][]{Park2014, Mace2018} on Gemini South (GS, 8.1 m) to observe the hot Jupiter WASP-77\,Ab (1740 K, 1.36 day period, 1.21\,R$_J$, 1.76\,M$_J$).  With a single (4.7 hrs) continuous observation of the pre-eclipse phases (0.32 $< \phi <$ 0.47, where $\phi=0$ is transit, and $\phi=0.5$ is secondary eclipse),  \cite{line2021} were able to obtain precise constraints on both the TP 
and on the volume mixing ratios of the dominant carbon and oxygen bearing gases, CO and H$_2$O ($\pm$0.1-0.2 dex).  These CO and H$_2$O constraints enabled precise metallicity (quantified by the sum of C and O relative to solar, $\pm$0.14 dex) and carbon-to-oxygen ratio constraints (C/O, $\pm$0.08), key diagnostics that link planetary atmospheres to their formation history.    

\begin{figure}
\begin{center}
\includegraphics[width=\linewidth]{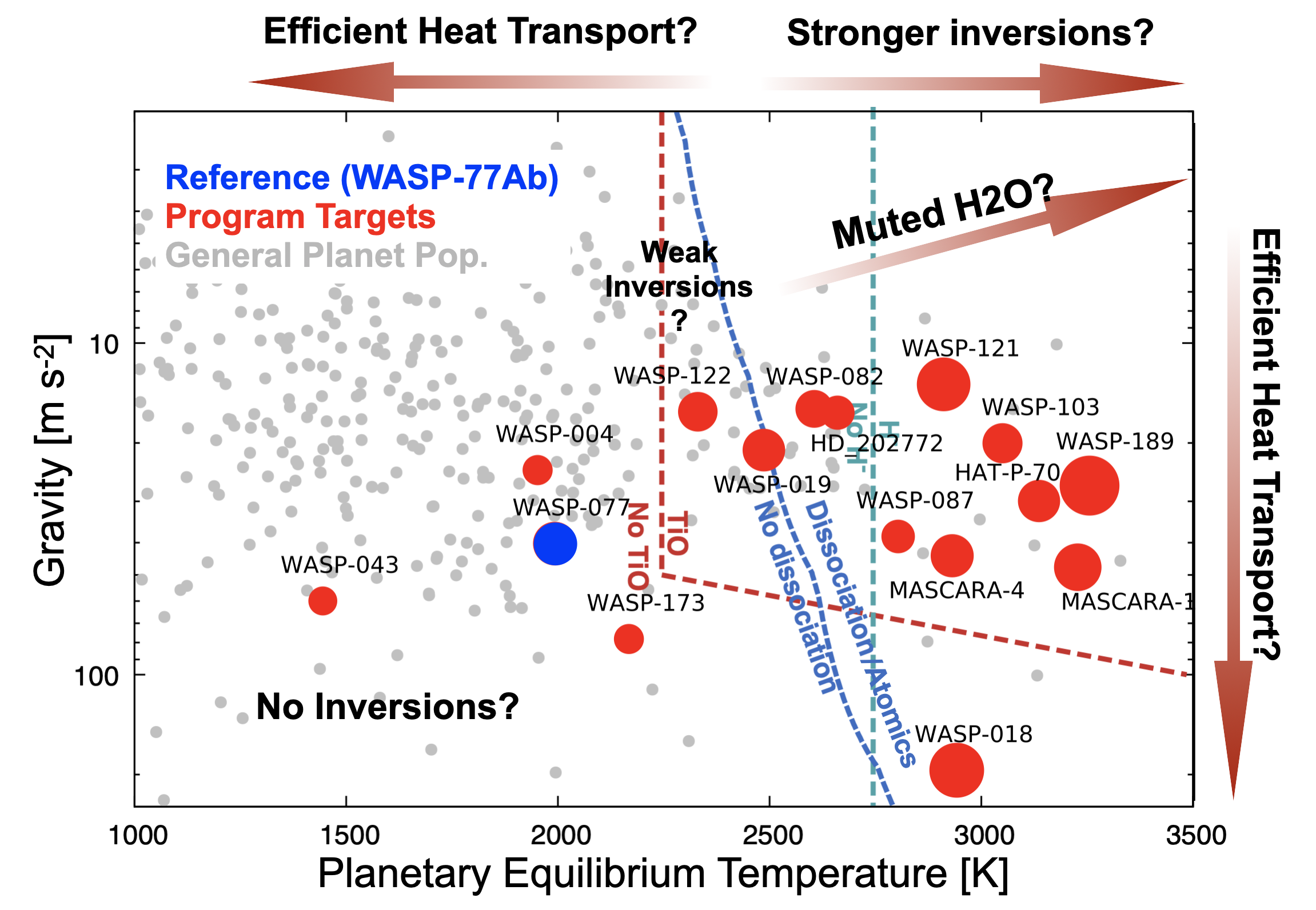}
\caption{Summary of the Roasting Marshmallows day side thermal emission HRCCS survey with IGRINS on Gemini South. The key goals are to identify trends in composition and thermal structure over a wide range of stellar irradiation (equilibrium temperature) and planetary mass (gravity).  The red circles are the proposed targets and blue is WASP-77\,Ab, recently observed with IGRINS in \cite{line2021}, for context. The symbol size is proportional to the relative emission signal-to-noise ratio for each target \citep[e.g.,][]{Kempton2017}, scaled to WASP-77\,Ab.  Qualitative temperature-gravity locations of key atmospheric transitions (onset of the major inversion causing species TiO/VO and the molecular thermal dissociation transition) are shown as dashed lines \citep[based upon][]{Parmentier2018}. Broad questions are also indicated in the appropriate regions of parameter space.   }
\label{fig:sample}
\end{center}
\end{figure}

Building upon the successful atmospheric characterization performance of IGRINS on GS shown in \cite{line2021},
we embark on an IGRINS survey (see Figure \ref{fig:sample}) to measure the thermal emission from the day side hemisphere of a population of over a dozen hot-to-ultra-hot Jupiters. The goals of the program are to 1) determine the vertical temperature profiles as a function of planetary gravity and stellar influx (equilibrium temperature) and 2) assess the dispersion in the intrinsic elemental inventory of C and O.   In this manuscript we present our first results of our survey for the ultra-hot Jupiter, WASP-18b \citep{Hellier2009}.

WASP-18b is a well studied \citep{Maxted2013, Wong2020} transiting ultra-hot Jupiter, owing to its high equilibrium temperature (2,400\,K), short orbital period (0.94 days), favorable transit depth (1.2\%, $R_\mathrm{P}$=1.191\,$R_\mathrm{J}$, $R_\mathrm{S}$=1.26\,$R_\odot$), and bright host star ($V$=9.27 mag, $K$=8.13 mag, $T_\mathrm{eff}$=6,400\,K, spectral type F7)\footnote{\url{https://www.astro.keele.ac.uk/jkt/tepcat/planets/WASP-018.html}}. The high gravity ($\log g$=4.35, $M_\mathrm{P}$=10.5\,$M_\mathrm{J}$) is more akin to those of brown dwarfs rather than ``typical" transiting planets, which may preclude it from transmission spectroscopy. However, numerous HST orbits and Spitzer hours have gone into characterizing its thermal emission either through secondary eclipse \citep{Nymeyer2011, Sheppard2017, Arcangeli2018} or phase curve spectro-photometry \citep{Maxted2013, Arcangeli2019}. The measured hot day side ($\sim$2,900\,K), relatively cool night side ($\lessapprox$1,500\,K), and small eastward phase curve peak offset ($\sim 5 ^{\circ}$) suggest relatively inefficient day-to-night heat transport, in line with predictions for planets in this temperature range \citep{Iro2005, Perez-Becker2013a}. 

More intriguing are the mixed conclusions regarding the atmospheric composition. \cite{Sheppard2017} analyzed secondary eclipse observations with HST WFC3-G141 and four Spitzer IRAC channels combined with an atmospheric temperature and abundance retrieval analysis \citep{Madhusudhan2011}. They found a high mixing ratio of CO ($\approx$10-40\%, driven by the Spitzer 4.5/3.6 $\mu$m flux ratio) and an upper limit on the water abundance (due to the featureless HST WFC3-G141 spectrum), leading to the inference of a high metallicity ($\approx$100-700$\times$ solar) and high C/O ($\approx1$). Such an extreme composition is largely at odds with expectations for the compositions of Jovian-like words \citep{Thorngren2016, Madhusudhan2014a}.  An independent analysis of the same data set \citep{Arcangeli2018}, but with a different modeling strategy, found an approximately solar composition atmosphere. The key differences between the analyses were primarily model driven.  The latter assumed a more physically self-consistent model that included the effects of both water thermal dissociation and H- bound- free/free-free continuum opacity, both of which, along with CO, are expected to sculpt the near-infrared spectra of ultra-hot Jupiters \citep{Arcangeli2018, parmentier18}.  Both works concluded that the dayside thermal structure was ``inverted" (rising temperature with altitude), as predicted \citep{Fortney2008, Hubeny2003}.  \cite{Gandhi2020} provided yet another independent modeling analysis of various versions of the aforementioned data sets and a hierarchy of minimal-assumption retrieval models to conclude that the published data are largely uninformative and possibly drive the composition towards un-physical solutions. 

Given the current atmospheric composition ambiguity of this keystone ultra-hot Jupiter, more observations are needed to disentangle the key physical processes and composition. Here, we present an independent analysis of this planet with a different but complementary set of data. Furthermore, WASP-18\,b's perplexing atmosphere and favorable observability is why it is included as one of three planets observed in the JWST Transiting Early Release Science Program \citep[secondary eclipse with NIRISS SOSS,][]{Bean2018}.  

Next we describe our observations and data reduction, followed by our molecular detections and atmospheric abundance and temperature retrieval results placed within the context of the aforementioned space-based observations. 

\section{Observations \& Initial Reduction} \label{sec:obs}
\begin{figure}[h]
\begin{center}
\includegraphics[width=0.5\textwidth]{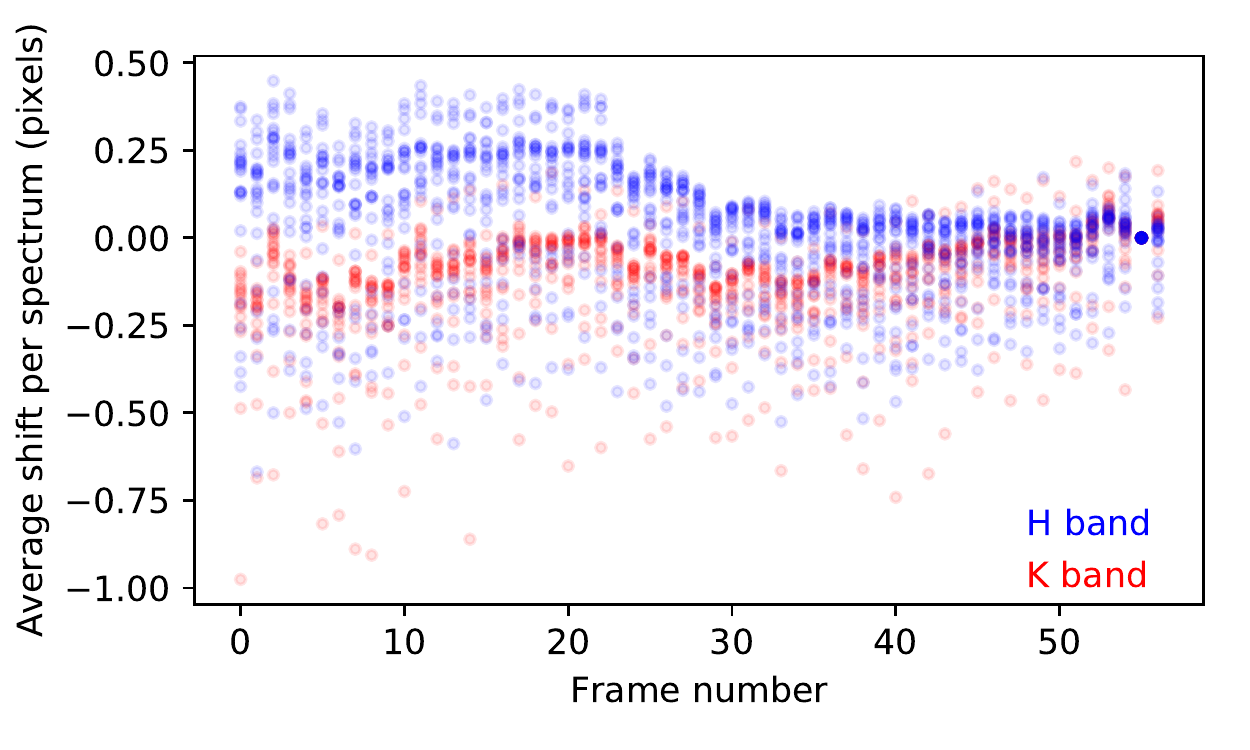}
\caption{Measured shifts (in pixels) versus time of the IGRINS spectral sequence for orders in the $H$ band (blue, one dot per order) and in the $K$ band (red dots).}
\label{fig:shifts}
\end{center}
\end{figure}

We observed WASP-18b on September 25, 2021 with IGRINS on Gemini South as part of the 117 hour Large-and-Long Program, ``Roasting Marshmallows: Disentangling Composition \& Climate in Hot Jupiter Atmospheres through High-Resolution Thermal Emission Cross-Correlation Spectroscopy" (GS-2021B-LP-206, PI M. Line, see Figure \ref{fig:sample}). A continuous sequence of 57 AB pairs (in an AB-BA nodding pattern, 100s exposure per AB-pair) were obtained over a course of 2.85 hours covering the pre-eclipse orbital phases ($0.326 < \phi < 0.452$), which are primarily sensitive to the day side hemisphere thermal emission. Here orbital phases were computed by using the time of mid-transit and orbital period provided in \citet{Shporer2019}. We achieved per-resolution element signal-to-noise ratios between 150 and 290 per AB pair depending on order, comparable to the WASP-77\,Ab observations in \citet{line2021}.
The initial data reduction (optimal extraction, wavelength calibration) is performed by the IGRINS instrument team using the IGRINS Pipeline Package \citep{Mace2018,LeePLP2016}. As described in \cite{line2021}, we perform an additional 
refinement of the pixel-wavelength solution to correct for the $\pm$0.2 pixel ($\pm$0.46 km s$^{-1}$) shifts measurable throughout the night (see Figure~\ref{fig:shifts}).
We discard 12 orders heavily contaminated by telluric lines and/or with very low instrumental throughput, namely the wavelength ranges $<1.44$ $\mu$m, 1.79-1.95 $\mu$m, and $>2.42$ $\mu$m. Furthermore, we remove the first and last 100 pixels from each order to discard spectral regions with negligible flux due to the instrumental blaze function. 

Following standard practice, the spectra are packaged into a $N_\mathrm{ord} \times N_\mathrm{int} \times N_\mathrm{pix}$ data cube for subsequent analyses, where $N_\mathrm{ord} = 42$ is the number of spectral orders, $N_\mathrm{int} = 57$ is the number of spectra, and $N_\mathrm{pix} = 1848$ the number of spectral channels after the order selection and edge trimming described above. 

In ground-based HRCCS observations, the dominant spectral features from telluric and stellar lines must be removed from the spectral sequence. 
As these contaminants are stationary (or quasi-stationary) in wavelength, it is possible to model them as time-correlated trends in common between spectral channels. 

Following \cite{DeKok2013} and \cite{line2021} we use the Singular Value Decompositon (SVD) method in the time domain to isolate the dominant trends (a.k.a. components) and subtract their best-fit linear combination to correct for the observed flux variations. We remove 5 components in this case, but the result is very weakly dependent on the number of components within the tested range (3-11 components). We apply three additional steps compared to \cite{line2021}:
\begin{itemize}
    \item For each order, we exclude spectral channels with flux levels less than 2\% of the median value computed across wavelengths. This step prevents the SVD algorithm to focus on decomposing noisy spectral channels where telluric lines are close to saturation. Depending on the order, we discard between 0 and 8.6\% of the data.
    \item For each order, we mask spectral channels with strong residuals \textsl{after} subtracting the SVD fit, with a dynamic threshold dependent on the number of spectral channels available per order, and ranging between 3.24 and 3.27$\times$ the mean standard deviation of the residuals\footnote{Given $N_\mathrm{ch}$ spectral channels, the threshold is computed as via \texttt{scipy.stats.norm.isf} as the inverse survival function of a normal distribution with argument $1/N_\mathrm{ch}$.}. The additional number of spectral channels masked is typically below 1\%, and at most 2.7\%. Over the 42 orders analyzed, 1.9\% of the data is excluded from subsequent steps due to the pre- and post-SVD masking.
    \item We apply a high-pass filter to the residuals, by convolving each spectrum with a Gaussian kernel of FWHM = 80 pixels, and subtracting the smoothed spectrum. This step removes any broad-band correlated noise along the wavelength axis (the $N_\mathrm{pix}$ dimension).
\end{itemize}

We note that in the literature \citep[e.g.,][]{Giacobbe2021} there are alternative versions of SVD, or more generally Principal Component Analysis (PCA), used to clean high-resolution spectra from telluric and stellar contaminants. In Appendix~\ref{sec:pca_vs_svd}, we show that the signal detected is robust against the choice of such de-trending algorithm, i.e. retrieved velocities and scaling parameters are consistent within 1$\sigma$. 

\begin{figure*}[!t]
\includegraphics[width=\textwidth]{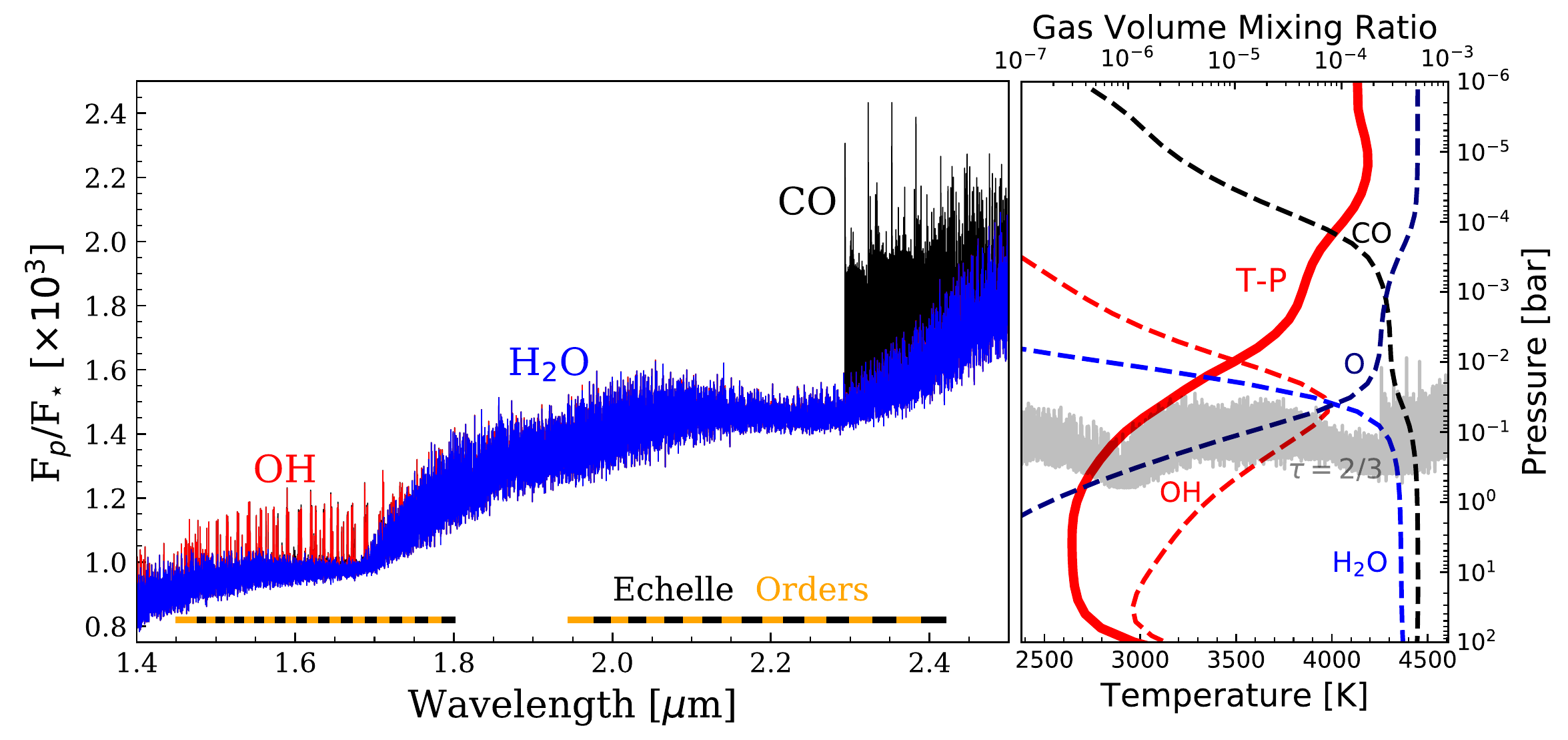}
\caption{Components of the spectrum that contribute to the observed signal. The template spectrum is a result of a 1$\times$ solar composition atmosphere in 1D radiative-convective-thermochemical equilibrium, assuming a redistribution factor of 2.35 \citep[based upon the trend in][]{Parmentier2021}. The dominant absorbers are H$_2$O, OH, and CO. Starting with H$_2$O (on top of an H- bound-free/free-free continuum), the other gases are sequentially added to illustrate their contribution to the total spectrum. The right panel shows the atmospheric structure (gas volume mixing ratio profiles for H$_2$O, OH, CO, and O as dashed lines and the temperature-pressure profile in solid red) from the 1D-RCTE model. The atomic oxygen (O) profile is also shown to illustrate its role in the total elemental oxygen inventory.  Also shown is the $\tau=2/3$ surface (in gray) to identify the pressure levels sensed by the observations--between 1 and 0.01 bar. This suggests that the observations are sensitive to the region where the mixing ratio profiles change rabidly due to thermal dissociation as a result of the rapidly increasing temperature with altitude.}
\label{fig:spectra_components}
\end{figure*}

\section{Molecular Detections with Cross-Correlation} \label{sec:CCF_analysis}

We perform an initial search for molecular absorbers in the atmosphere of WASP-18\,b by comparing our data to template spectra derived from planet specific 1D radiative-convective-thermochemical  equilibrium (1D-RCTE) models \citep{pis18,Arcangeli2018, Mansfield2021} at solar composition  post processed at an $R$=250,000 and subsequently convolved to a Gaussian instrumental profile corresponding to an $R$=45,000 (Figure \ref{fig:spectra_components}). We look for the signatures of H$_2$O, CO, and OH as they are the primary species expected to be present over the IGRINS pass-band for any reasonable atmospheric compositions. We also search for the more exotic species Ca, Si, and FeH.  

\subsection{Detections in cross correlation}\label{sec:cc_maps}

\begin{figure}[!h]
\begin{center}
\includegraphics[width=0.45\textwidth]{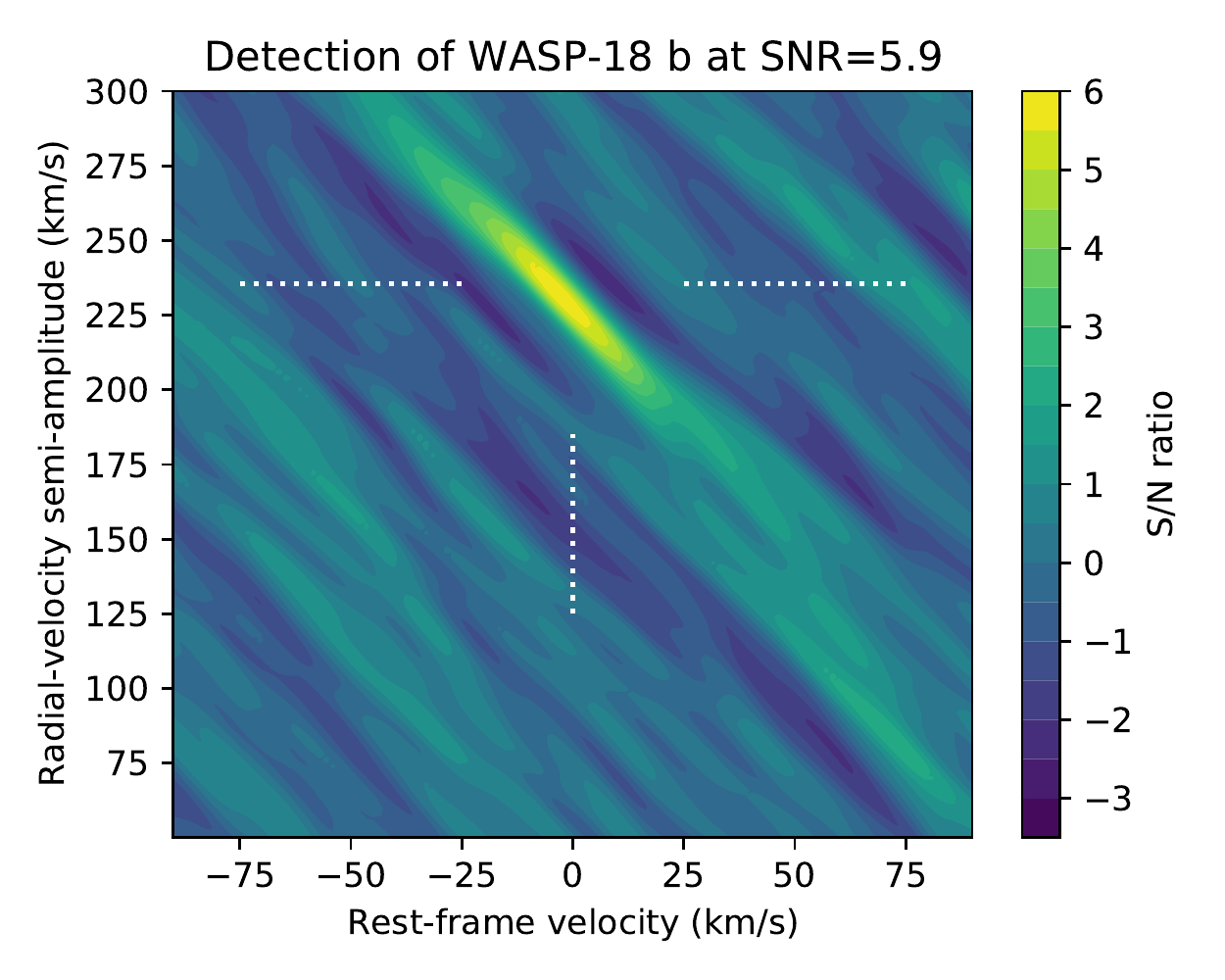}
\caption{Detection of the thermal emission spectrum of WASP-18\,b when cross correlating with the 1D RCTE model described in \S\ref{sec:cc_maps}. The planet is detected at a signal-to-noise ratio of 5.9, and at a position in velocity space consistent with the literature solution for the orbit and the velocity of the system (white dotted lines).}
\label{fig:fullBlob}
\end{center}
\end{figure}

We obtain a clear detection of the atmosphere of WASP-18\,b when cross correlating with the 1D-RCTE model including all (H$_2$O+CO+OH) the gases, at a signal-to-noise ratio of SNR=5.9. This is shown in Figure~\ref{fig:fullBlob} as a function of rest-frame velocity and planet maximum radial velocity. The dotted lines indicate the predicted planet position using the literature orbital solution from \citet{Shporer2019} and a systemic velocity of $V_\mathrm{sys} = 4.1\pm0.6$ km s$^{-1}$ \citep{GaiaDR2}. As standard in the literature, the quoted SNR is computed by dividing the peak cross correlation by the standard deviation of the noise away from the peak. To mitigate the somewhat subjective choice of noise samples, we fit the distribution of cross correlation values with a Gaussian profile and use the Gaussian $\sigma$ as proxy for noise.

\begin{figure*}[h]
\begin{center}
\includegraphics[width=\textwidth]{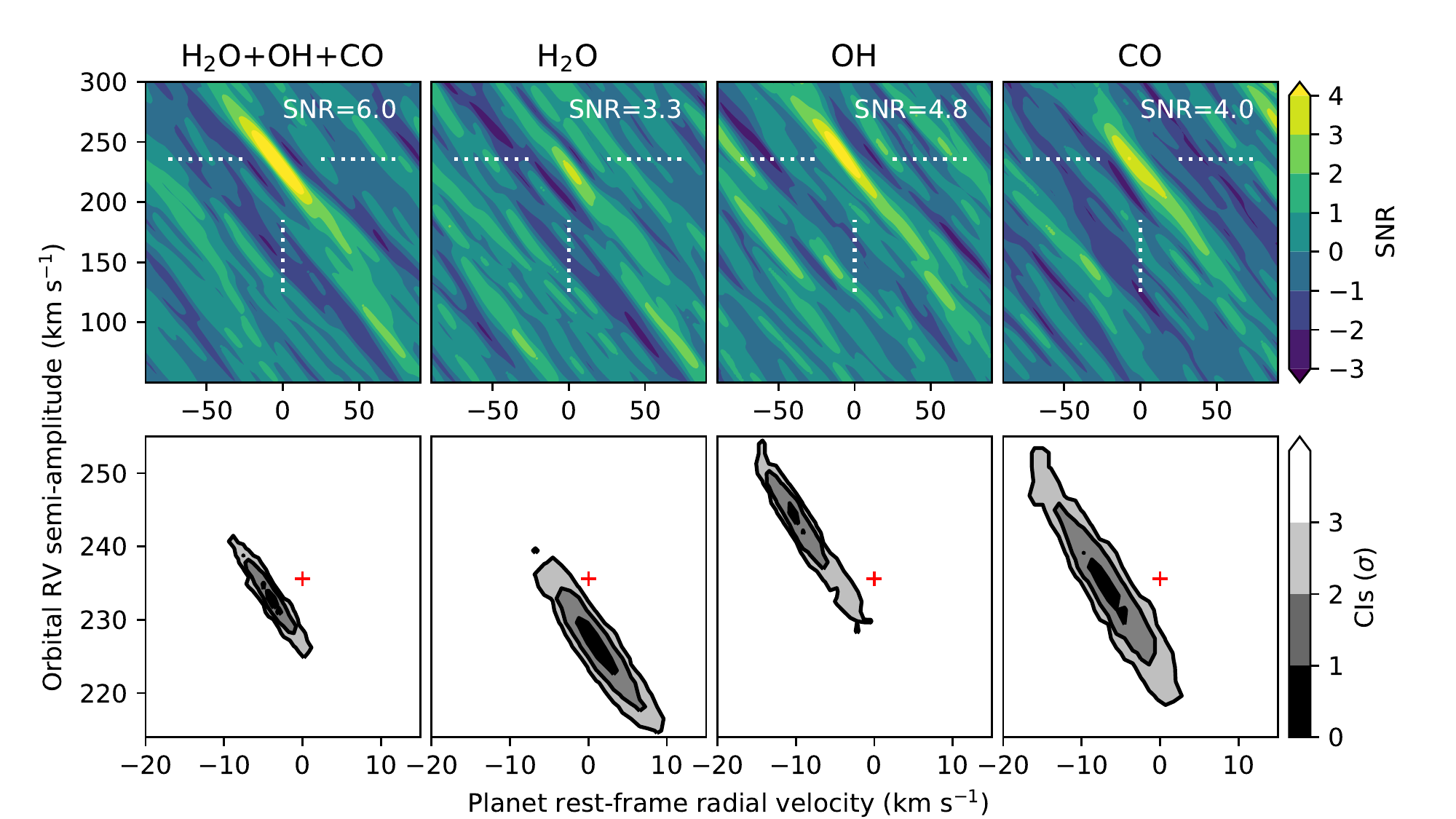}
\caption{\textsl{Top row:} Cross correlation maps of the species  detected in the atmosphere of WASP-18\,b using the model spectra shown in Figure \ref{fig:spectra_components}. Signal-to-noise ratios are computed by dividing the peak value by the standard deviation away from the peak (\S\ref{sec:cc_maps}). The literature solution for the planet velocity is indicated with white dashed lines. \textsl{Bottom row:} Confidence intervals for the same models as the top row, obtained via the likelihood framework and a 3-parameters MCMC as described in \S\ref{sec:lh_maps}. Each species is detected above the 3$\sigma$ level. Visual comparison of the offset contours to the literature solution (red cross)
indicates potential atmospheric dynamics as a function of species (\S\ref{sec:disc_vel_shift}).}
\label{fig:blobPlots}
\end{center}
\end{figure*}

Figure \ref{fig:blobPlots} (top panel) summarizes our search for each species via cross-correlation. We tentatively detect H$_2$O (SNR=3.3) and we obtain firmer detections of OH (SNR=4.8) and CO (SNR=4.0), the latter by restricting the analysis to the 4 orders of the $K$-band long-ward of 2.29 $\mu$m where the 2-0 ro-vibrational band-head of CO begins. When combined together, the three detected species are measured with a SNR=6.0. 

We do not detect neutral Ca, Si, or FeH individually. However, these species do not significantly lower the detection significance when included in the analysis (Figure \ref{fig:fullBlob}). We also note that the 1D RCTE models tested all possess a thermal inversion layer, and thus the positive correlation seen in Figures \ref{fig:fullBlob} and \ref{fig:blobPlots} suggests that an inverted atmosphere is indeed consistent with these data.

\subsection{Detections with the likelihood framework}\label{sec:lh_maps}

In Figure \ref{fig:blobPlots} (bottom panel), we show the velocity confidence intervals (CIs) derived through the likelihood framework of \citet{Brogi2019} for the same set of models presented in \S\ref{sec:cc_maps}. In order to properly estimate velocity CIs in such framework, it is important to match the overall line depth in addition to their position, because changes in line depth produce a stretch of the log-likelihood function. Therefore, we add a global log-scaling factor $\log(a)$ in addition to the two velocities and sample the 3-dimensional posteriors via the Python \texttt{emcee} package, with priors listed in the top three rows of Table \ref{tab:priors}, and with the atmospheric spectra fixed to the 1D-RCTE spectrum described in \S\ref{sec:CCF_analysis}. When using the likelihood framework, it is imperative to reproduce the effects of the data analysis on each tested model. Such \textit{model reprocessing} is implemented by multiplying the SVD fit obtained in \S\ref{sec:obs} by the appropriately Doppler shifted, scaled model, and by reapplying the SVD algorithm on the injected model sequence with the same number of components and mask as for the observed data. In this analysis, applying a high-pass filter to the reprocessed model does not noticeably change derived CIs, and therefore we skip this step to reduce computational times.

While detection significance is not measurable via likelihood ratio, we can look at the drop in significance between the best-fit velocities and the rest of the posterior as a proxy for it. For the three species detected in \S\ref{sec:cc_maps}, the peak likelihood is favoured at more than 4$\sigma$ over any other signal, demonstrating the superior performance of the likelihood framework compared to pure cross correlation in localising the signature of molecular species in velocity-amplitude space. This is in line with the expectation that the likelihood of \citet{Brogi2019} uses not only the information about the line position, but also about the line shape and depth. Despite the higher sensitivity, no bound posterior is obtained when testing the models containing Ca, Si, and FeH. 

The results from the likelihood framework are suggestive of different dynamics probed by the three detected species. This effect is visible in Figure \ref{fig:blobPlots} as a shift in the best-fit rest-frame velocities for H$_2$O, CO, and OH. We further discuss these velocity shifts in \S\ref{sec:discussion}.

\section{Temperature \& Abundance Retrieval Analysis} \label{sec:retrieval}
The above search for molecular species provides qualitative insights into the atmospheric composition.  In order to obtain quantitative constraints on the molecular abundances and temperature profile (and subsequent products like the atmospheric metallicity and C/O ratio) we must perform an atmospheric retrieval. Following \citet{line2021}, \citet{kasper2021}, and \cite{Kasper2022}, we applied the \citet{Brogi2019} cross-correlation-to-log-likelihood retrieval framework to derive the molecular volume mixing ratios and the vertical temperature structure.  While going through the retrieval process we gleaned critical insights into performance of two HRCCS retrieval frameworks: ``free-retrievals" and ``1D-RCTE grid retrievals" (``gridtrieval"). The processes of interpreting the free-retrieval results in $\S$\ref{sec:discussion} caused us to question the applicability of free-retrievals to UHJ's. This then led us to employ the 1D-RCTE based ``gridtrieval" method described in $\S$\ref{sec:grid_retrieval}.  Below, and in the following sections, we document our path chronologically, the assumptions we tested, and ultimately the results we settle upon. 

We begin the retrieval process with the basic \texttt{CHIMERA} ``free-retrieval" \citep{Line2013a,kreidberg2015} paradigm which assumes constant-with-altitude gas mixing ratios and uses the 3-parameter \citet{Guillot2010} parameterization for the TP profile. We start with this method/paradigm as it worked well for the cooler planet, WASP-77Ab, in \cite{line2021}. The retrieval parameters specific to our analysis here and their prior ranges are given in Table \ref{tab:priors}. A more detailed description of the high-resolution GPU-based radiative transfer method, including opacity sources\footnote{Transition and broadening information is taken from a variety of sources and is used to pre-compute cross-sections on a $T$-$p$-wavenumber grid. The H$_2$O \citep{Polyansky2018} and FeH \citep{Bernath2020MOLLIST} data are sourced from the \hyperlink{https://www.exomol.com/data/molecules/}{ExoMol project} \citep{Tennyson2020} and the cross-sections generated using the methods described in \citet{GharibNezhad21}. We use \hyperlink{https://helios-k.readthedocs.io/en/latest/}{HELIOS-K} \citep{Grimm2021} to generate the OH and CO cross-sections drawn from the \hyperlink{https://hitran.org/hitemp/}{HITEMP} database \citep{hitemp2010, gordon2022hitran2020}. HCN \citep{BarberHCN, Tennyson2020} cross-sections are also generated with HELIOS-K.  Finally, we use the \citet{kurucz1995}  \hyperlink{https://lweb.cfa.harvard.edu/amp/ampdata/kurucz23/sekur.html}{atomic line information} to compute the Ca+ cross-sections using a custom routine and the H- free-free/bound-free cross-sections are generated using the methods described in \citet{John1988}. }, and log-likelihood implementation within {\tt pymultinest} \citep{feroz2009,Buchner2014} is given in \cite{line2021}. 

\begin{figure*}
\begin{center}
\includegraphics[width=\textwidth]{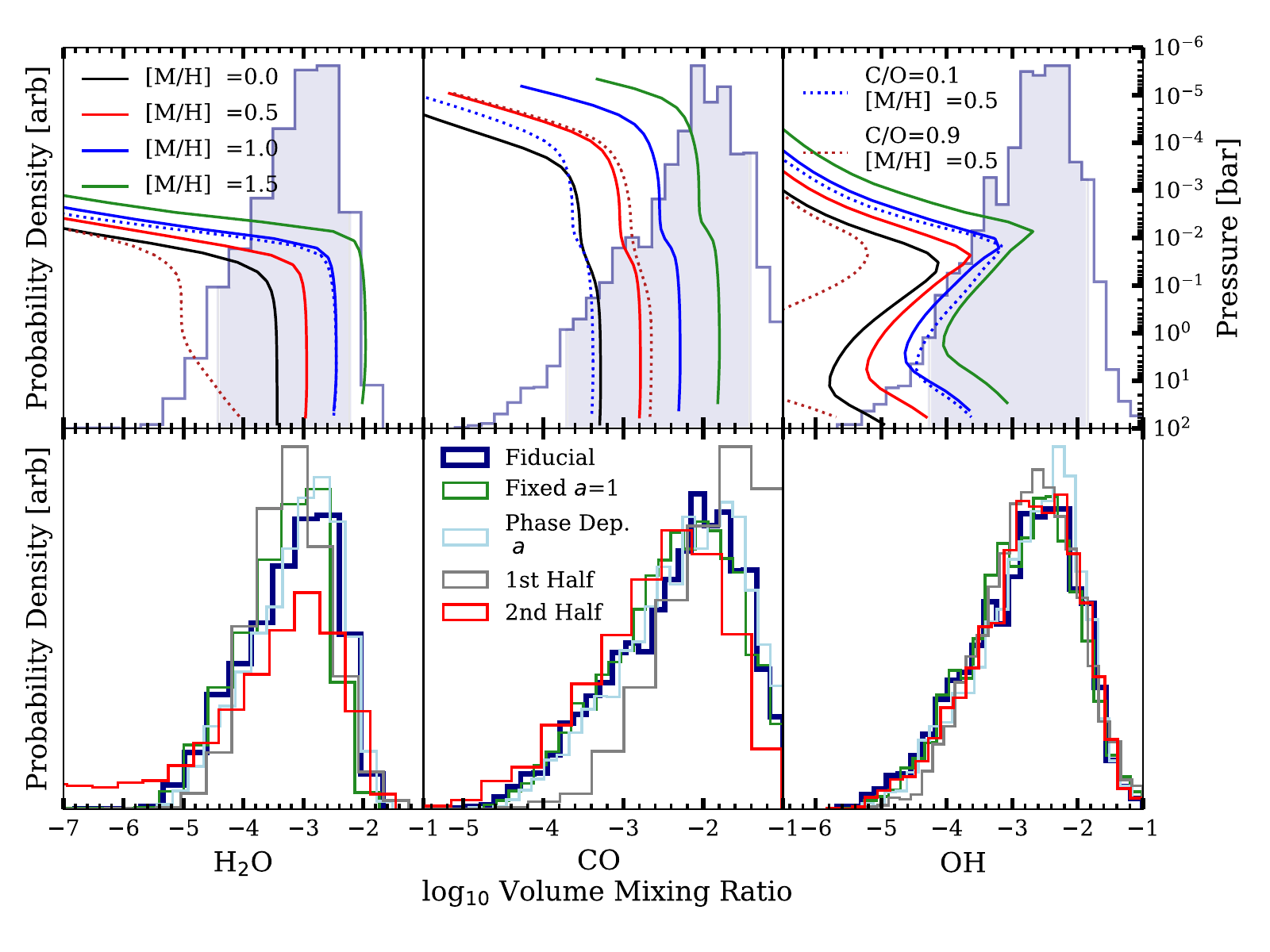}
\caption{Summary of the ``free retrieval" results for the constrained species, H$_2$O, CO, and OH. \textsl{Top row:} Marginalized gas mixing ratio histograms (95\% confidence interval shown as shaded region)  under the ``Fiducial" scenario described in $\S$\ref{sec:fiducial} compared to a series of variable composition self-consistent 1D-RCTE model abundance profiles (see footnotes 4-6).  The retrieved abundances are consistent with a wide range of metallicities, but their combination can rule out ``self-consistent" combinations of gases at high C/O.  \textsl{Bottom row:} Summarizes the impact of the free retrieval assumptions on the marginalized gas mixing ratio histograms (see $\S$\ref{sec:testing_loga} and \S\ref{sec:phase_resolved}). These assumptions appear to have a minimal quantitative impact on the gas volume mixing ratios.}
\label{fig:AbundHist}
\end{center}
\end{figure*}

Within the free retrieval paradigm, we perform a series of exploratory experiments with various assumptions in order to test the robustness of the results. Figure \ref{fig:AbundHist} and Figure \ref{fig:TP} summarize these results compared with a battery of WASP-18b specific 1D-RCTE\footnote{For 1D-RCTE models one has to assume a ``heat redistribution" \citep{Fortney2005a} to derive the mean ``day side" hemispheric properties--a variable that multiplies the incident stellar flux.  We define redistribution as $(T_\mathrm{day}/T_\mathrm{eq})^4$ with $T_\mathrm{day}$ the ``dayside temperature" and $T_\mathrm{eq}$ the equilibrium temperature. A value of unity implies  $T_\mathrm{day}=T_\mathrm{eq}$, a value of two implies  $T_\mathrm{day}=2^{1/4}T_\mathrm{eq}$. The maximum physical value is 2.67 \citep{Cowan2011, Arcangeli2019}} models \citep{Arcangeli2018, Mansfield2021} of different compositions. Table \ref{tab:posteriors} gives the numerical values for 68\% confidence intervals for each parameter under each scenario. The full posteriors for each scenario are given as corner plots in Appendix \ref{sec:app_corner_plots}.

\begin{deluxetable}{lcl}
\tablecaption{\label{tab:priors} Description of retrieved parameters and their prior ranges. All priors are assumed uniform between the bounds given. Variables correspond to the labeling in the corner plot shown in the Appendix.}
\tablehead{
 \colhead{Parameter}  & \colhead{Description} &  \colhead{Prior} 
}
\startdata
$K_\mathrm{P}$ & Planet Orbital velocity &  200 -- 300\,km\,s$^{-1}$\\
$V_\mathrm{rest}$ & Excess rest-frame velocity  & -30 -- 30\,km\,s$^{-1}$ \\
log($a$) & model multiplicative  & -1 -- 1\\
& scale factor & \\
log($\gamma_1$) & vis-to-IR opacity & -3 -- 3 \\ 
log($\kappa_{\text{IR}}$) & IR opacity & -3 -- 0.5 (cgs) \\
T$_{\text{irr}}$\tablenotemark{a} & irradiation temperature & 1500 -- 4500\,K\\
 H$_2$O, CO, OH & log gas volume & -12 -- 0 \\
 FeH, HCN, Ca I & mixing ratios & \\
 H- & for bound-free cont. & -15 -- -1 \\
 H*e$^-$  & for free-free cont.& -20 -- -1 \\
\enddata
\tablenotetext{a}{Irradiation temperature is really a measure of the equivalent {\it dayside} temperature. }

\end{deluxetable}		

\subsection{Fiducial retrieval}\label{sec:fiducial}
\begin{figure}[h]
\begin{center}
\includegraphics[width=0.45\textwidth]{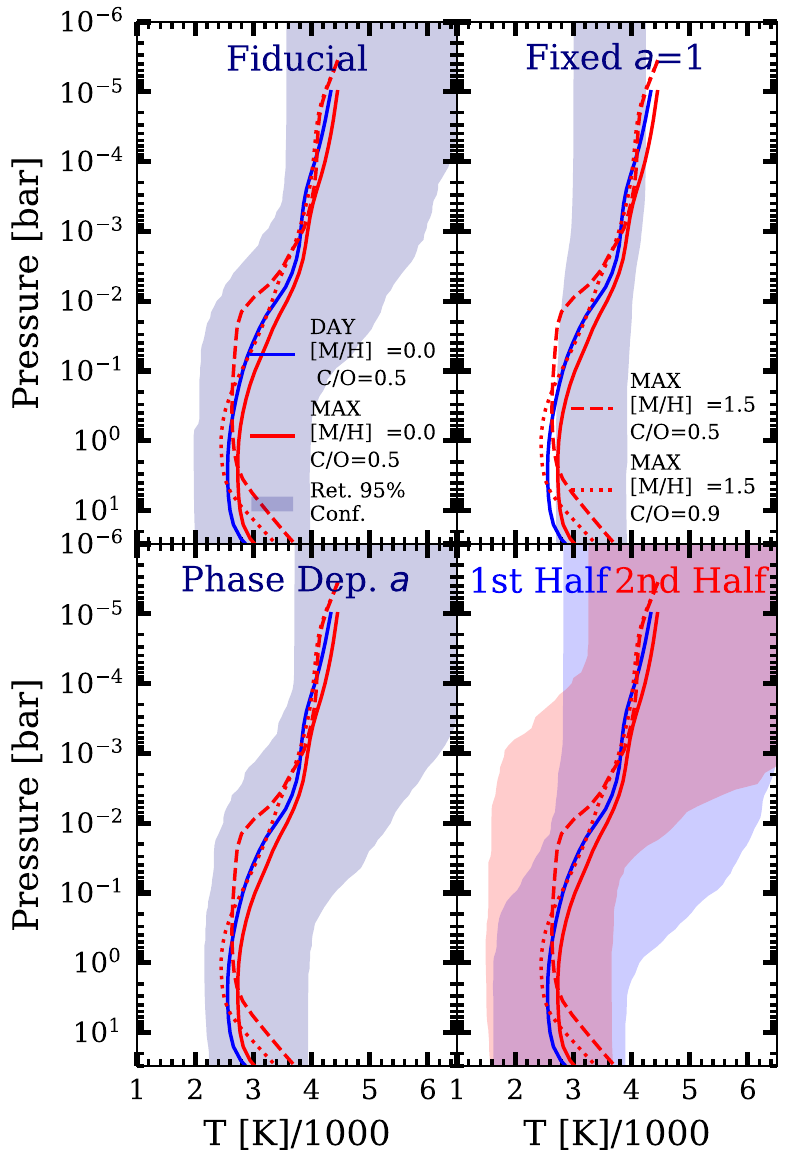}
\caption{Summary of the retrieved TP profiles and their sensitivity to the various free retrieval assumption scenarios described in $\S$\ref{sec:retrieval}, indicated at the top. These are summarized with the 95\% confidence interval derived from individual temperature-profiles reconstructed from 500 random posterior draws (based upon the $T_\mathrm{irr}$, log($\kappa_\mathrm{IR}$), and log($\gamma_\mathrm{v}$) constraints).  We also include select self-consistent 1D-RCTE TP profiles to facilitate comparison.  Here, DAY represents a redistributing scaling factor of 2 and MAX is 2.67 (see footnote 3).  Despite the wide confidence interval, the general morphology of the retrieved TP profiles is consistent with the 1D-RCTE predictions (e.g., rising temperature with decreasing pressure).      }
\label{fig:TP}
\end{center}
\end{figure}

Under the fiducial setup (``Fiducial" in Figures \ref{fig:AbundHist}, \ref{fig:TP}, and in Table \ref{tab:posteriors}), we retrieve for all of the parameters with the given priors shown in Table \ref{tab:priors} on the full data sequence.  We obtain bounded constraints on the volume mixing ratios of H$_2$O, CO, and OH and upper limits on all of the other species we explored (Table~\ref{tab:posteriors}). These results are consistent with the detections in the cross-correlation analysis (\S\ref{sec:CCF_analysis}). As in \citet{line2021}, we attempt to constrain the $^{13}$CO/$^{12}$CO isotopologue ratio, but arguably due to the lower signal to noise of the detection we can only place an uninformative upper limit on their ratio (see Appendix Figure \ref{fig:corner_full_15pars}).  
The top panel of Figure \ref{fig:AbundHist} shows the gas mixing ratio constraints (95\% confidence intervals shown with the shaded region) on H$_2$O, CO, and OH from this setup compared to those predicted by series of 1D-RCTE\footnote{Here, we assume a redistribution factor of 2.2, consistent with the ``nightside clouds" trend in \cite{Parmentier2021}. The C/O is assumed to be 0.5 unless otherwise stated.} models of differing compositions (via metallicity\footnote{We use the \cite{Lodders2009} elemental abundances  as our solar reference. [M/H]=0, 1, 2 represent 1, 10, 100$\times$ the solar elemental values, respectively} [M/H], and C/O).  The retrieved mixing ratios are consistent with physically plausible combinations of these species and with metallicities between $\sim$solar ([M/H]=0.0) and $<$30$\times$ solar ([M/H]=1.5).  The constraints on water rule out the higher metallicities, whereas the constraints on OH tend to prefer metallicities elevated above solar.  The CO constraints are consistent with a wide range of metallicities.  We also show the impact of the elemental C/O on the 1D-RCTE profiles (as dashed lines) under the [M/H]=0.5.  The ``high" C/O (0.9) scenario is in tension with both the water and OH abundances. Low C/O values (0.1) have little influence on the 1D-RCTE H$_2$O and OH abundance profiles, but do impact CO directly. In $\S$\ref{sec:discussion} we discuss a more detailed elemental abundance analysis. Here we simply aimed to illustrate that regions of the retrieved gas mixing ratios and their combinations  are chemically plausible under a wide range of elemental compositions. 

These data also provide coarse constraints on the vertical temperature profile. Specifically,  the visible-to-IR opacity, log($\gamma_v$), which governs the atmospheric temperature gradient; values of log($\gamma_v$)$<$0 produce decreasing temperatures with decreasing pressure, 0 produces isothermal atmospheres, and $>$0 result in temperatures that increase with decreasing pressure, or inversions. Our retrieved median log($\gamma_v$) is larger than zero by more than $3\sigma$ (see Appendix Figure \ref{fig:corner_full_15pars}), a clear indication of a temperature structure with a thermal inversion. 

We obtain only lower limits on the retrieved irradiation temperature ($T_\mathrm{irr}$), with most of the posterior above the planet's equilibrium temperature of $T_\mathrm{eq}\approx2,400$\,K. As defined in footnote 4, we can relate the retrieved irradiation temperature back to a ``heat redistribution" efficiency, assuming that the retrieved irradiation temperature represents the {\it dayside} temperature.  The maximum redistribution efficiency of 2.67 would only allow dayside temperatures up to 3,065 K for WASP-18b. With the posterior pushing to the edge of the prior at 4,500 K, our fiducial model thus retrieves nonphysical temperatures for this planet, which we believe is due to the known correlation with the scaling factor $a$. 

The retrieved log-scaling factor is not bounded in this Fiducial retrieval run, which might indicate the inability of the model to reproduce both the shape and the depth of spectral lines in the presence of thermal dissociation. With a low retrieved log-scaling factor, the line amplitude has to be compensated by an increase in the planet irradiation temperature and/or in the lapse rate.

We also observe a degeneracy between the log($\kappa_\mathrm{IR}$) parameter, which controls the ``vertical pressure shift" in the TP profile via the mapping between optical depth ($\tau$) and pressure ($\tau=\kappa_\mathrm{IR}P/g$), and gas mixing ratios. As pointed out in past works \citep{Line2015, Piette2020}, increasing the abundances of a gas shifts the ``average" $\tau=2/3$ surface to lower pressures. To maintain a constant temperature gradient over the line formation pressures, the whole TP profile must ``shift up" by increasing $\kappa_\mathrm{IR}$.  
While pressure broadening alters the line shape and should thus help ``anchoring'' the TP profile at high-enough SNR, at first order or low SNR the above degeneracy is only halted when either a gas abundance or $\kappa_\mathrm{IR}$ run up against their prior edge, as can be seen in Appendix Figure \ref{fig:corner_full_15pars}.  

The 95\% confidence band resulting from the reconstructed TP profiles, marginalized over all of the aforementioned degeneracies, is shown in the upper left panel (``Fiducial") in Figure \ref{fig:TP}.  The uncertainties are about 1,000\,K, much larger than those obtained for our previous target, WASP-77\,Ab.

We test the sensitivity of absolute abundances to the scaling factor in \S\ref{sec:testing_loga}, while we discuss the shortcomings of the model and propose the alternative retrieval (``gridtrieval") scheme in \S\ref{sec:disc_chemistry} and \S\ref{sec:grid_retrieval}, respectively.

\begin{deluxetable*}{lcccccc}
\tablecaption{\label{tab:posteriors} Retrieved parameters for the five retrieval setups described in \S\ref{sec:retrieval}. Quoted uncertainties are 1$\sigma$ intervals for the bounded parameters, 2$\sigma$ for upper or lower limits. Dashes denote unconstrained parameters.}
\tablehead{
 \colhead{Parameter} & \colhead{Units}  & \colhead{Fiducial} &  \colhead{1$^\mathrm{st}$ half} & \colhead{2$^\mathrm{nd}$ half} & \colhead{Phase Dep. $a$} & \colhead{Fixed $a$=1} 
}
\startdata
log$_{10}n_\mathrm{H2O}$ & - & $-3.13^{+0.66}_{-0.98}$ & $-3.21^{+0.62}_{-0.64}$ & $< -2.01$ & $-3.03^{+0.62}_{-0.92}$ & $-3.28^{+0.62}_{-0.89}$ \\
log$_{10}n_\mathrm{CO}$ & - & $> -4.12$ & $> -3.36$ & $> -4.50$ & $> -3.91$ & $> -4.04$ \\
log$_{10}n_\mathrm{Ca}$ & - & $< -5.13$ & $< -3.96$ & $< -6.67$ & $< -5.02$ & $< -4.54$ \\
log$_{10}n_\mathrm{FeH}$ & - & $< -6.66$ & $< -5.98$ & $< -6.58$ & $< -6.66$ & $< -6.62$ \\
log$_{10}n_\mathrm{HCN}$ & - & $< -2.04$ & $< -2.29$ & - & $< -1.89$ & $< -2.23$ \\
log$_{10}n_\mathrm{OH}$ & - & $> -4.74$ & $> -4.20$ & $> -4.61$ & $-2.63^{+0.63}_{-0.94}$ & $> -4.55$ \\
$[^{13}$CO/$^{12}$CO$]$ & - & $< 1.70$ & $< 2.12$ & $< 1.24$ & $< 1.64$ & $< 1.58$ \\
log$_{10}n_\mathrm{H-}$ & - & $< -6.62$ & $< -7.25$ & $< -5.06$ & $< -6.48$ & $< -6.66$ \\
log$_{10}n_\mathrm{H}n_{e-}$ & - & - & $< -2.00$ & - & - & - \\
log$_{10}(\gamma_v)$ & - & $0.74^{+0.40}_{-0.36}$ & $0.74^{+0.55}_{-0.40}$ & $> 0.45$ & $0.86^{+0.36}_{-0.36}$ & $0.17^{+0.06}_{-0.04}$ \\
log$_{10}(\kappa_\mathrm{IR})$ & - & $> -2.45$ & $< -0.30$ & $> -2.28$ & $> -2.31$ & $> -2.22$ \\
T$_\mathrm{irr}$ & K & $> 2086$ & $> 1791$ & - & $> 2187$ & $3559^{+ 288}_{- 391}$ \\
$K_\mathrm{P}$ & km s$^{-1}$ & $234.6^{+ 2.9}_{- 2.9}$ & $232.6^{+ 6.3}_{- 7.4}$ & $244.3^{+ 6.7}_{- 5.8}$ & $235.1^{+ 3.0}_{- 3.0}$ & $234.5^{+ 2.9}_{- 2.7}$ \\
$V_\mathrm{rest}$ & km s$^{-1}$ & $-7.2^{+2.6}_{-2.3}$ & $-5.2^{+5.9}_{-5.2}$ & $-11.7^{+3.2}_{-3.2}$ & $-7.7^{+2.6}_{-2.0}$ & $-4.9^{+1.9}_{-1.9}$ \\
log$_{10}(a)$ & - & $< -0.10$ & $< 0.20$ & $< -0.05$ & $< -0.21$ & - \\
$\Delta\ln\mathcal{Z}$\tablenotemark{a} & - & 0 & N/A & N/A & 0.59 & 1.34 \\
\enddata
\tablenotetext{a}{Fiducial minus the given scenario, e.g., $\ln\mathcal{Z}_{Fid.}-\ln\mathcal{Z}$ }
\end{deluxetable*}

\subsection{Phase-dependent and fixed scaling factor}\label{sec:testing_loga}
As the daysides of UHJ's are anticipated to be much hotter than the nightsides \citep[e.g.,][]{Arcangeli2019}, it is not unreasonable to presume that any phase dependency in flux is simply due to a phase dependent dayside dilution. We observe the more irradiated, hence hotter, dayside hemisphere closer to secondary eclipse, which progressively becomes ``area" diluted by a cooler, lower flux nightside (which we take to be negligible), at phases closer to quadrature. We parameterize this effect with a phase-dependent scale factor, $\log[a(\varphi)]$, modulated with cosine function \citep{Burrows2006} which is 0 at $\varphi= 0$ (mid-transit, only night side visible), and 1 at $\varphi= 0.5$ (mid-eclipse, full dayside) (``Phase Dep. $a$" in Figures \ref{fig:AbundHist}, \ref{fig:TP}, and in Table \ref{tab:posteriors} ), though we still retrieve for a constant multiplicative factor, $a$, on top of this phase dependent scaling.  Upon doing so, we do not see any meaningful difference in either the abundances (Figure \ref{fig:AbundHist}, bottom row) or temperature (Figure \ref{fig:TP}, bottom left) with the Fiducial retrieval. The Bayesian evidence difference (bottom row, Table \ref{tab:posteriors}) between this and the Fiducial scenarios are considered insignificant on a Jeffery's scale \citep{trotta2008bayes}, confirming the lack of meaningful effect.

Lastly, we fix $\log(a) = 0$ (``$a=1$" in Figures \ref{fig:AbundHist}, \ref{fig:TP}, and in Table \ref{tab:posteriors}) to test for any dependence of the measured abundances on the scaling factor. In \S\ref{sec:fiducial} we showed that the retrieved scaling factor has a posterior leaning towards the lower edge of the prior. For ultra-hot Jupiters we would expect scale factors below unity as the spectral shape we observe is dominated by a locally hotter region on the dayside hemisphere, but the overall flux is decreased due to the ``area" dilution from the cooler off-hot-spot regions of the dayside hemisphere. 
However, a posterior strongly leaning towards $\log(a) = -1$ would imply a reduction of a factor of 10 in the strength of emission lines. Given that at the orbital phases probed the sub-stellar point is still in view and the emission predominantly comes from the planet's dayside, such a small value of $a$ cannot be explained with the dilution argument above. It is therefore worth testing whether such a small scaling factor can impact abundance measurements. By looking at Table~\ref{tab:posteriors} and the bottom row of Figure \ref{fig:AbundHist}, it is clear that the abundances retrieved with $\log(a) = 0$ are identical to those from the fiducial run within a small fraction of the 1-$\sigma$ CI. Inspecting the other parameters reveals that the change in scaling factor is completely absorbed by the TP profile, which is much shallower in this case to compensate for the larger amplitude of spectral lines. 

In the next section, we explore the impact of the phase range--which maps to longitude on the planet--on the retrieved properties.


\subsection{Phase-resolved retrieval}\label{sec:phase_resolved}
In order to explore potential longitudinal variations in composition, temperature, and winds, we split the spectral sequence in half and separately retrieve on each sub-set. The ``1st half" covers $0.326 < \phi \le 0.389$ and the ``2nd half" covers $0.389 < \phi \le 0.452$. The latter half covers more of the irradiated dayside than the former. We show the full posteriors of these runs in the Appendix Figures \ref{fig:corner_1st_half} and \ref{fig:corner_2nd_half}.

First we focus on the changes on the abundance of the three detected species (H$_2$O, CO, OH), shown in the bottom row of Figure~\ref{fig:AbundHist}. 
These changes are negligible at most, and can be summarized in a weaker detection of water for the ``2nd half" (when more of the hot-spot is visible) and a very marginal increase in the CO abundance for the ``1st half". We do not see any notable shift in the chemistry of the other undetected species. The bottom right panel in Figure \ref{fig:TP} compares the reconstructed TP profiles retrieved from each half-sequence.  The large uncertainties preclude the identification of any obvious differences, with perhaps a minor shift in pressure. The log($\gamma_v$) constraints (Table \ref{tab:posteriors}) from both half-sequences indicate the obvious presence of an inversion. 

Finally, we look for shifts in the retrieved $K_\mathrm{P}$ and $V_\mathrm{rest}$ as proxies for possible atmospheric dynamics. We retrieve an excess $V_\mathrm{rest}$ shifting from $-5.2^{+5.9}_{-5.2}$ km s$^{-1}$ for the first half to $-11.7 \pm 3.2$ km s$^{-1}$ for the second half, with the fiducial retrieval sitting somewhat in the middle ($V_\mathrm{rest} = -7.2^{+2.6}_{-2.3}$ km s$^{-1}$). However, we note that $V_\mathrm{rest}$ and $K_\mathrm{P}$ are highly correlated parameters in emission spectroscopy when just the pre- or post-secondary eclipse phases are observed. In fact, when we explore the two-dimensional posterior in both $V_\mathrm{rest}$ and $K_\mathrm{P}$, the solutions from the first and second halves are still consistent at $\sim1\sigma$. 
We briefly note here that phase effects measured via HRCCS are different from phase curves measured at low spectral resolution or multi-band photometry. Whereas the latter mostly probe changes in the planet's continuum, the former encode changes in the line depth, i.e. they map the temperature difference between the continuum and the line cores. Therefore, we should not necessarily expect HRCCS phase curves to match low-resolution phase curves, particularly when investigating shifts of the maximum amplitude from the sub-stellar point \citep[e.g.,][]{vansluijs22}.

\section{Discussion}\label{sec:discussion}

\subsection{Deriving the chemistry of WASP-18\,b from retrieved abundances}\label{sec:disc_chemistry}

Compared to our pilot study for WASP-77 Ab \citep{line2021}, the WASP-18\,b retrieval struggles to constrain the absolute abundances of the three detected species. In fact, the measured precision of 1 dex or better is being influenced by the prior limit on $\kappa_\mathrm{IR}$ (e.g., see the log$_{10}(\kappa_{IR})$ row in Figure \ref{fig:corner_full_15pars}), rather than being entirely constrained by the data alone.
Nevertheless, the correlation between species visible in Appendix Figures \ref{fig:corner_full_15pars} to \ref{fig:corner_full_14pars} suggests a higher precision in measuring {\it relative} abundances, which should help constraining elemental abundance ratios such as C/O.

Here we infer the ``free-retrieval" based C/O and metallicity of WASP-18\,b by counting the total elemental number density arising from each species (Table ~\ref{tab:posteriors}, column ``Fiducial''). We then compare the measured chemistry to that of the parent star. Importantly, we do not assume solar composition for WASP-18. As mentioned in footnote 5 we take the solar elemental abundances from \citet{Lodders2009} (C/O$_\odot$=0.46). For the star WASP-18, we use the recent abundance study of \citet{Polanski2022}, where abundances are relative to the solar values of \citet{Grevesse2007} (Alex Polanski, private communication)--similar but not identical to \citet{Lodders2009}. Remarkably, the reported stellar C/O in WASP-18 is 0.23$\pm$0.05\footnote{We acknowledge the challenges in obtaining stellar C/O ratios \citep{Fortney2012,Brewer16}.  A model independent analysis \citep{Bedell2018} of solar neighborhood solar twins suggests that C/O's $<$0.4 are unlikely. Furthermore, \cite{KoleckiWang22} find systematically higher C/O's than \cite{Polanski2022} for overlapping stars (though they do not include WASP-18\,b in their analysis). }, i.e. significantly lower than the solar value. The stellar metallicity is approximately solar ([M/H] = 0.045).

To compute the planetary C/O and metallicity with the above in mind, we use the following formulae:
\begin{equation}\label{eq:c-to-o}
    \mbox{C/O} = \frac{n_\mathrm{C}}{n_\mathrm{O}} =
    \frac{n_\mathrm{CO} + n_\mathrm{HCN}}{n_\mathrm{CO}+n_\mathrm{OH}+n_\mathrm{H2O}},
\end{equation}
and
\begin{equation}\label{eq:metall}
[\mbox{M/H}] = \log\left\{\frac{2n_\mathrm{CO}+2n_\mathrm{HCN}+n_\mathrm{OH}+n_\mathrm{H2O}}{2 n_\mathrm{H2}\,[(n_\mathrm{O}+n_\mathrm{C}+n_\mathrm{N})/n_\mathrm{H}]_\odot}\right\},
\end{equation}
where the numerator in Equation \ref{eq:metall} sums over the number densities of all the species containing metals (``M") relative to the assumed hydrogen fraction ($n_\mathrm{H}$=2$n_\mathrm{H2}$) in the free retrieval\footnote{In the free retrieval, the filler gas is assumed to be H$_2$+He (with $n_\mathrm{He}$/$n_\mathrm{H2}$=0.176). Certainly this assumption breaks down in UHJs where atomic H dominates. As long as the total H is accounted for, this assumption should not matter for elemental counting purposes. }. We thus note that $n_\mathrm{H2}$ in the denominator is different for each posterior sample because it depends on the number densities of all the other gases. For reference, its median value is $n_\mathrm{H2}$=0.844.  The planetary ``M/H" is further normalized to the sum of the solar  C/H, O/H, and N/H. We find that the addition of Fe and Ca from the non-detected species makes a negligible difference in the measured planet metallicity, but they are still included in the computation of $n_\mathrm{H2}$.

\begin{figure}[h]
\includegraphics[width=0.43\textwidth]{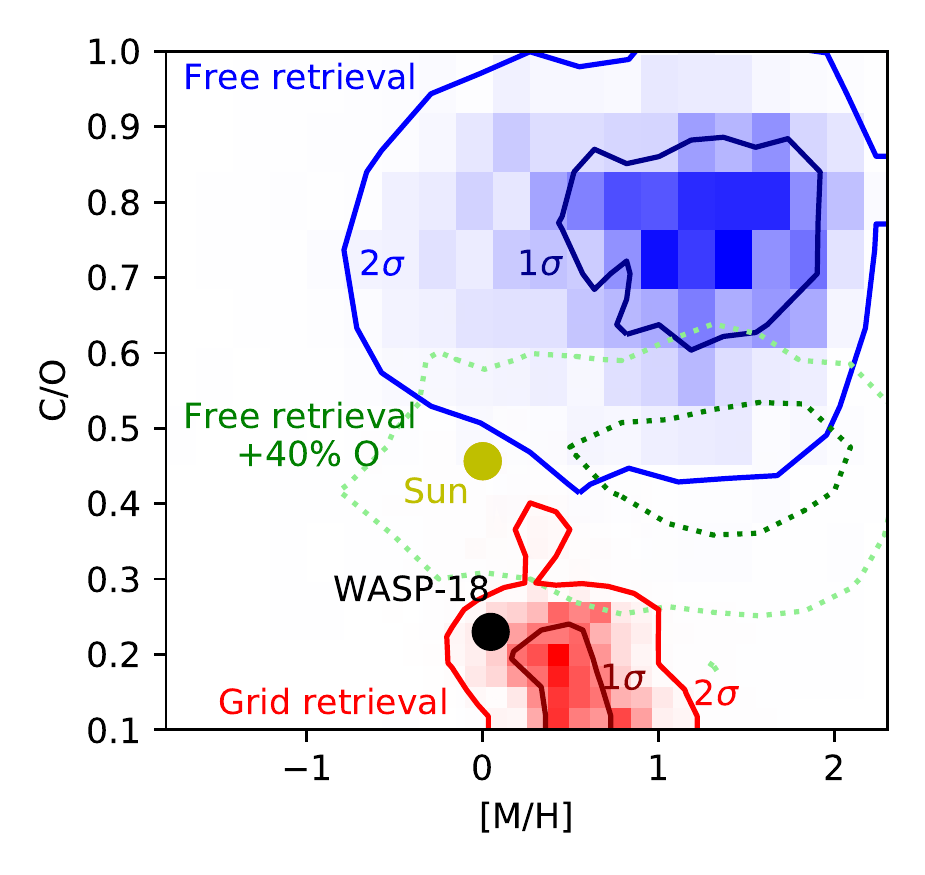}
\caption{Elemental abundances derived from the ``free retrieval" gas mixing ratio constraints (blue) versus the corresponding constraints from the self-consistent, grid-based 1D-RCTE retrieval (red). A correction to the free retrieval accounting for $\approx40\%$ atomic oxygen (non-detectable) is shown with the green dotted lines. Elemental values for the star, WASP-18, are shown as the black dot \citep{Polanski2022} and those for the Sun as the yellow dot \citep{Lodders2009}. The metallicity axis is normalized to the solar values (see Equation \ref{eq:metall}). }
\label{fig:co_metall}
\end{figure}

Figure~\ref{fig:co_metall} shows the joint probability for C/O and metallicity.  We obtain C/O$=0.75^{+0.14}_{-0.17}$ and [M/H]$ = 1.03^{+0.65}_{-1.01}$. The C/O is constrained to be less than 1 at the 2$\sigma$ level but is still markedly super-stellar. The metallicity is also super-stellar, although consistent with stellar values at 2$\sigma$. Overall, our free retrieval derived elemental abundances are indicative of a planetary atmosphere that is super-stellar in both metallicity and C/O at  $>$2$\sigma$.

There are, however, several shortcomings within the ``free-chemistry'' retrieval paradigm that might lead to biases--especially in UHJs. Firstly, the strong detection of OH (and correspondingly weaker detection of H$_2$O) is an indirect confirmation of thermal dissociation. Thermal dissociation will convert H$_2$O into both OH and atomic O, and these infrared IGRINS observations are not sensitive to the latter. Unaccounted atomic O ($\ge$40\% of the total O inventory at $p < 10^{-2}$  bar--Figure \ref{fig:spectra_components}, right) could  potentially produce an upwards bias of any C/O estimates based on Equation~\ref{eq:c-to-o} alone. If the measured $n_\mathrm{O}$ is scaled upward by 40\%, we indeed obtain a lower C/O$=0.45^{+0.08}_{-0.10}$ and a slightly higher [M/H]$ = 1.17^{+0.66}_{-1.01}$ (Figure \ref{fig:co_metall}, green dotted lines). Even with this correction, the chemistry derived from the free retrieval is incompatible with the parent star at $\ge$2$\sigma$.

Furthermore, with dissociation processes one should expect strongly altitude dependent gas mixing ratio profiles \citep{parmentier18, lothringer18}. In particular, the abundance of H$_2$O rapidly decreases at the atmospheric pressures where dissociation begins, with corresponding increase in the OH abundance, which is indeed visible in the 1D-RCTE models shown in Figures \ref{fig:spectra_components} and \ref{fig:AbundHist}. To counter-balance this effect,   the constant-with-altitude retrieval models assumed here will attempt to compensate by adjusting the absolute abundances, changing the lapse rate (through the $\gamma_V$ parameter), or shift the overall TP profile (through the $\kappa_\mathrm{IR}$ parameter). These adjustments defy simple intuition, due to the known correlation between abundances and lapse rate, as well as to the saturation of strong CO and OH line cores forming in the upper isothermal layer. 

Finally, all free retrievals must assume a ``filler gas", e.g., H$_2$O, OH, and CO are all trace gases and the remaining is some mixture of H$_2$, He, and atomic H but we don't know the relative proportions as we have little sensitivity to the hydrogen induced continua.  All in all, we suspect that absolute abundances from the ``free-chemistry'' retrieval might be biased due to the shortcomings of our current modelling, impacting the accuracy of the elemental abundance estimates. We note, however, that these free retrieval assumptions are valid for the WASP-77\,Ab analysis in \citet{line2021} as that planet lies within a more suitable temperature regime (e.g., no thermal dissociation, relatively constant with altitude abundances for the dominant C and O species).

In the following section, we quantitatively address these issues by building a modeling framework capable of self-consistently computing the TP and gas volume mixing ratio profiles through a grid of 1D-RCTE models. 

\subsection{Self-consistent Grid-based Retrieval: ``Gridtrieval"}\label{sec:grid_retrieval}
\begin{figure*}
\begin{center}
\includegraphics[width=\textwidth]{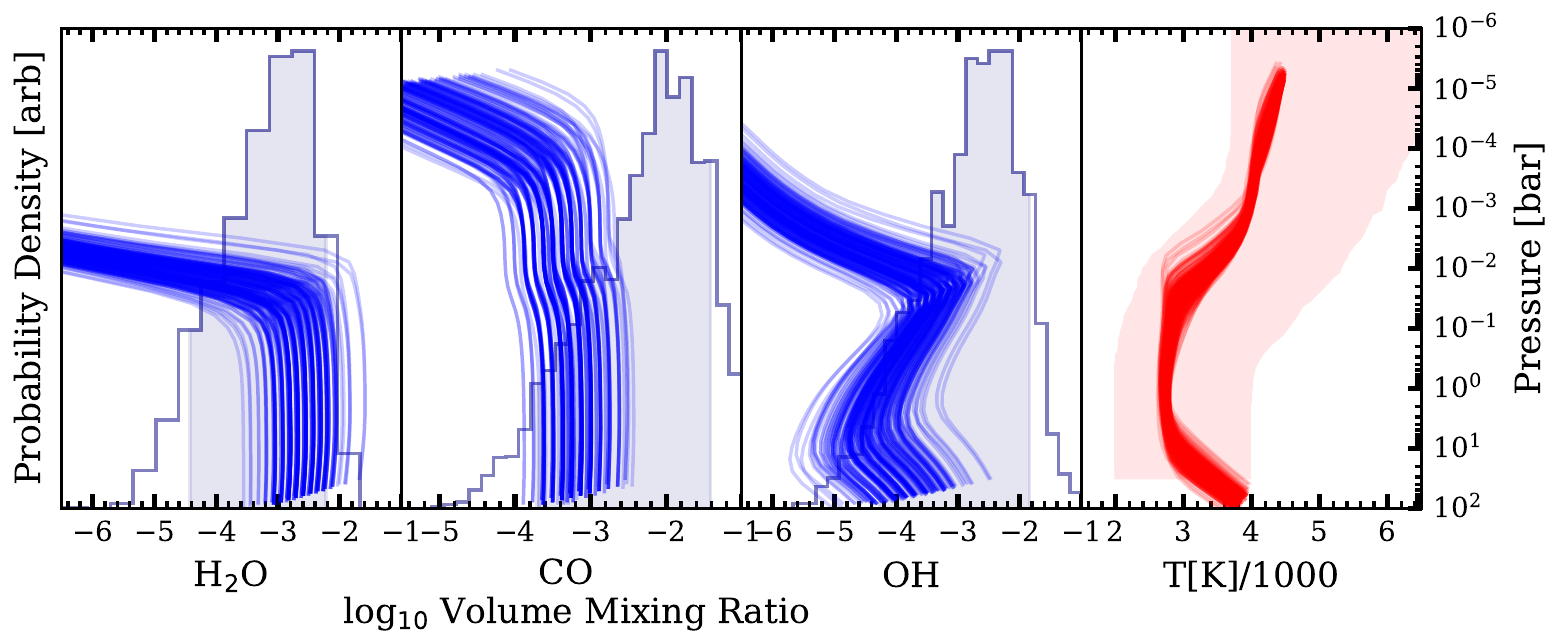}
\caption{Grid-based retrieval gas volume mixing ratios and temperature profiles compared to the Fiducial free retrieval constraints.  The grid-based 1D-RCTE atmospheric gas volume mixing ratio- (left three panels in blue) and temperature- (right most panel in red) pressure profiles (solid curves) are drawn from the grid retrieval posterior samples.  The free retrieval histograms and 95\% confidence intervals for H2O, CO, and OH (left 3 panels)  are the same as those in the top panel of Figure \ref{fig:AbundHist}. The right most panel shows the 95\% confidence interval of the reconstructed free retrieval TP profiles in light red (same as the top left panel in Figure \ref{fig:TP}). There is general agreement in both the volume mixing ratios and temperature-pressure profiles.    
  }
\label{fig:AbundHist_Grid}
\end{center}
\end{figure*}

Instead of retrieving for the individual gas volume mixing ratios and a parameterized temperature profile, we fit directly for a redistribution, metallicity, and C/O derived from 1D-RCTE model fits. As a reminder, within the 1D-RCTE models, the temperature and gas vertical mixing profiles are self-consistently computed given the elemental abundances and stellar irradiation.  To do this, we first build a grid of 5,600 WASP-18b 1D-RCTE models as a function of redistribution (1.85-2.65, equivalent of irradiation temperatures of 2,800-3,060 K in 20 K steps), [M/H] ($-1.0$ to 2.0, 0.125 dex increments), and C/O (0.1-0.95, 14 points with with variable spacing). Here, metallicity (or rather 10$^{[M/H]}$) is a rescaling factor to the elemental abundances relative to H \citep{Lodders2009} and C/O adjusts the relative C and O while preserving their sum {\it after} adjusting the total metallicity.  We then replace our free retrieval forward model (all other aspects remain the same) with a grid of high resolution spectra (with the relevant opacities) generated from the 1D-RCTE model atmospheres and fit for these 3 ``grid" parameters, along with the usual velocities and scale factor. A nearest neighbor search is used to identify the closest [redistribution, [M/H], C/O] point for a given {\tt pymultinest} live point parameter set.  

The results of this analysis are shown in Figure \ref{fig:corner_grid_retrieval} in Appendix \ref{sec:app_corner_plots} (which also illustrates the grid point spacing), and the posteriors in C/O and [M/H] in Figure \ref{fig:co_metall}. This method gives a C/O$<$0.34 (2$\sigma$ upper limit) and [M/H]$ =0.48^{+0.33}_{-0.29}$. The redistribution parameter constraints run up against the prior upper limit value of 2.65 (effectively the maximum physically allowed value).  Similar to free retrieval derived temperatures, these data are unable to place a bounded constraint on the absolute temperature information. However, in contrast to the free retrievals, we do not see a noticeable degeneracy between temperature information and composition (perhaps a slight correlation between C/O and redistribution). 

Inspecting Figure \ref{fig:co_metall} we find that there is no overlap in the elemental abundance constraints derived from the free and 1D-RCTE grid retrievals at better than the 2$\sigma$ level. While the marginalized metallicities are consistent (0.48$^{+0.33}_{-0.29}$ vs. 1.03$^{+0.65}_{-1.01}$), the C/O's are vastly different, with the grid retrieval preferring C/O ratios below $\sim$0.3 and the free retrieval indicating C/O ratios $>$0.5. Correcting (green contours in Figure \ref{fig:co_metall}) for the ``missing O" in the free retrievals helps some, but does not completely remedy the disagreement.   Figure \ref{fig:AbundHist_Grid} elucidates an additional potential cause of these discrepancies. While the ensemble of reconstructed 1D-RCTE atmospheric profiles generally falls within the 95\% confidence intervals of the Fiducial free retrieval constraints, the latter contains more flexible composition combinations as the free retrievals are not bound by 1D-RCTE physio-chemical constraints.  

The elevated C/O in the free retrievals is being driven by the elevated CO abundances (in equation \ref{eq:c-to-o}) of which are not possible over the prescribed 1D-RCTE grid dimensions and physical constraints. In a UHJ, equilibrium chemistry along an inverted TP profile results in a CO/H$_2$O ratio that increases with decreasing pressure due to thermal H$_2$O dissociation. The constant-with-altitude assumption within the free retrieval prohibits this behavior. The 1D-RCTE cannot reproduce this high CO abundance parameter space.  For instance, the 1D-RCTE CO abundance profile can be increased to the upper edge of the free retrieval confidence interval by increasing  [M/H] to 2. However, this also increases the H$_2$O abundance profile beyond the free retrieval confidence interval.  Increasing the C/O to 0.9 can dial back the H$_2$O abundance, however that also forces the CO abundance to go beyond the free retrieval confidence interval while also lowering the OH profile to fall below its free retrieval confidence interval.  This game can be played with varying combinations of [M/H] and C/O, but the resulting profiles struggle to agree with the free retrieval confidence intervals as well as those that were indeed found by grid retrieval. 

Finally, the Bayesian evidence difference between the grid retrieval and Fiducial free retrieval scenario ($\Delta$ln$Z$=$-$1.9) is considered weak-to-moderate, suggesting that there is no strong preference for one method over the other (albeit, a slight preference for the grid-based retrieval owing to the vastly reduced prior volume), meaning that the grid based retrieval is not over-constrained relative to the free retrieval. 

In summary, the grid retrieval constraints on redistribution, [M/H], and C/O result in temperature and vertical mixing ratio profiles that fall within the free retrieval confidence intervals--they are consistent. The restricted mixing ratio possibility space resulting from the imposed self-consistency from radiative-convective-thermochemical equilibrium simply reduces the possible combinations of H$_2$O, CO, and OH. Relaxing 1D-RCTE assumptions in the free-retrievals permits higher values of CO that result in higher C/O ratios.

\subsection{Are molecules tracing different dynamics?}\label{sec:disc_vel_shift}

\begin{figure}[h]
\begin{center}
\includegraphics[width=0.45\textwidth]{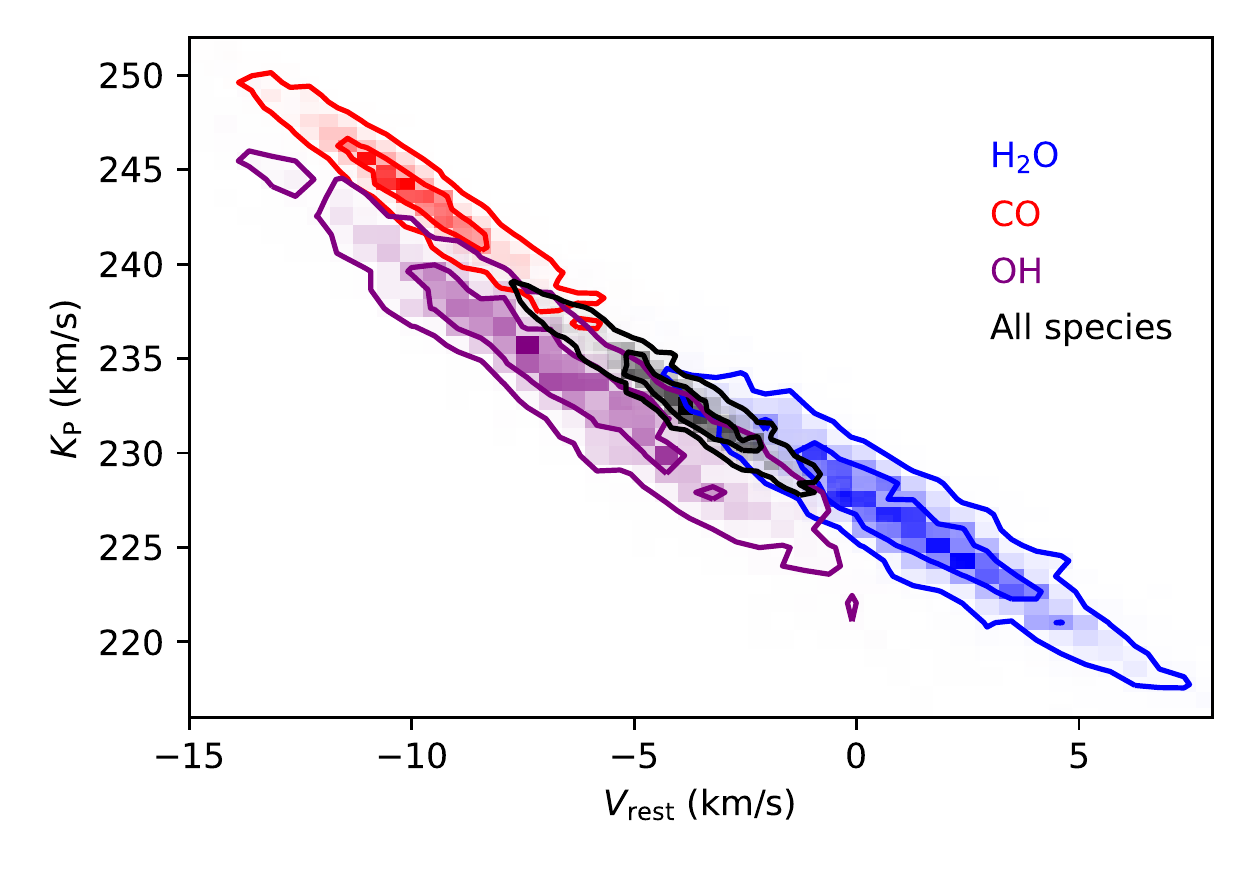}
\caption{1- and 2-$\sigma$ confidence intervals for H$_2$O (blue), CO (red), OH (purple), and their mixed spectrum (black) in rest-frame velocity and planet orbital velocity. It shows a blue shift detected for CO and OH, marginally ($\sim$2$\sigma$) inconsistent with H$_2$O. Further discussion in \S\ref{sec:disc_vel_shift}.}
\label{fig:vel_shifts}
\end{center}
\end{figure}

The single-molecule analysis presented in \S\ref{sec:CCF_analysis} reveals a tantalising shift between the best-fit velocity solution from water vapour and those from CO and OH. We plot in Figure~\ref{fig:vel_shifts} the 1- and 2-$\sigma$ velocity confidence intervals from the three species. While water vapour is approximately centered around the expected rest-frame velocity, both OH and CO appear marginally shifted. 

Given the final precision in the wavelength calibration of our data (better than 200 m s$^{-1}$ for all the spectral orders), we exclude that the measured shifts are due to an imprecise instrumental calibration. Furthermore, \citet{Gandhi2020xsec} found that the line lists of CO and H$_2$O are appropriate for high resolution studies up to $R$=100000, which excludes shifts due to inaccurate line positions. Larger line-list uncertainties for other species such as CH$_4$ \citep[evident, e.g., in the T dwarf spectra of][]{Tannock2022} might instead impact similar velocity analyses. In our case, we also note that shifts are measured in both $K_\mathrm{P}$ and $V_\mathrm{rest}$, whereas line-list inaccuracies should produce a global shift, i.e., a shift in $V_\mathrm{rest}$ only. 

Doppler shifts in both $K_\mathrm{P}$ and $V_\mathrm{rest}$ point instead to a departure from the local ``slope'' of the Keplerian radial-velocity curve, which is more easily explained by invoking atmospheric circulation or heterogeneity effects. These could be due to different \textsl{altitudes} probed by different species, to their emission arising from different \textsl{locations} on the planet disk, or to a combination of both. 

In our analysis, the most evident shift is between CO and H$_2$O. \citet{Beltz2022} modelled the effects of magnetic drag and three-dimensional atmospheric circulation for WASP-76\,b, another ultra-hot Jupiter. They indeed predict a differential radial velocity between two wavelength regions dominated by water (1.2 $\mu$m) and CO (2.3 $\mu$m). However, around $\varphi$=0.4, their water appears blue-shifted compared to CO, which is the opposite of our findings. Furthermore, the magnitude of their Doppler shifts is modest ($\le$2 km s$^{-1}$) compared to our observations.

Focusing next on the milder velocity shift between OH and H$_2$O, this can be qualitatively explained if thermal dissociation (producing OH) mostly happens near the sub-stellar point, which is rotating \textsl{towards} the observer in the phase range probed by these observations (net blue shift). Water recombines near the day-night terminator in this picture, approximately facing the observer in the observed phase range and thus showing a near-zero excess radial velocity. Building a minimal quantitative model of the above effect requires projecting the planet's synchronous equatorial rotation of 6.45 km s$^{-1}$ along the line of sight of the observer, as well as adopting ad-hoc assumptions for atmospheric circulation patterns and spatial heterogeneity in the distribution of species. We argue that such model is beyond the scope of the present study and leave it to future work.

Turning towards the evidence of similar effects in the literature, \citet{cont21} reported a velocity shift between TiO and \ion{Fe}{1} from emission spectroscopy of WASP-33\,b. To explain the shift, the authors tentatively invoke a TiO-depleted hot spot, but they also caution against the authenticity of the TiO signal, due to previously inconclusive searches. \citet{sanchez-lopez2022} also report tentative shifts for different molecular species, this time in the infrared transmission spectrum of WASP-76~b and therefore probing the dynamics along the terminator rather than the planet's day side. Further evidence for species-dependent dynamics starts building up when including optical transmission spectroscopy. \citet{Kesseli2022} detected a dozen different atomic species in the atmosphere of WASP-76\,b and measured different velocity shifts in the $K_\mathrm{P}$-$V_\mathrm{rest}$ plane. Given their high SNR, they also identified asymmetries in the time-resolved cross-correlation function, an effect that was first discovered for \ion{Fe}{1} \citep{Ehrenreich2020}. Other searches for atoms in WASP-76\,b \citep{Tabernero2021} and WASP-189\,b \citep{prinoth22} also reported different shifts for different species, albeit limited to departures from the planet's $V_\mathrm{rest}$. Lastly, \citet{Stangret20} measured different velocity solutions for \ion{Fe}{1} and \ion{Fe}{2} in MASCARA-2\,b (KELT-20\,b), a result also obtained for the Balmer series, \ion{Na}{1}, and \ion{Ca}{2} \citep{Casasayas19}. 



Given the diversity of orbital phases probed (transmission versus emission spectroscopy), molecular and atomic species detected, and different irradiation regimes, it is still too soon to build a unified picture from the sparse measurements listed above. Nevertheless, the very substantial shifts of several km\,s$^{-1}$ measured in WASP-18\,b are potentially concerning from a modelling standpoint. The framework used here does not yet include the possibility to assign different Doppler signatures to different species, and this limit might lower the overall goodness of fit (i.e. the log-likelihood) of any mixed model if the true underlying spectrum has complex dynamics. One could even picture a worst-case scenario in which the retrieval, having to lock on only one velocity solution, would select that of the species with the strongest signature, thus artificially disfavouring the others. The above concern is another strong reason to still search for each species individually as done in \S\ref{sec:CCF_analysis}, even with the current availability of Bayesian retrievals. In these WASP-18\,b observations, it seems that mixed models are still capable of capturing three species at once with a common velocity solution (Figure \ref{fig:vel_shifts}), which is somewhat comforting. 


\subsection{Comparison to Past HST+Spitzer Constraints}\label{sec:vsHST}
As discussed in the introduction, \cite{Sheppard2017} concluded based upon their ``free retrieval" analysis on the HST WFC3+Spitzer IRAC data, that the C/O was unity, the metallicity (via C/H), was 145-680$\times$ Solar, and that the temperature profile possessed a strong inversion. They argue that this solution is driven by the lack of a distinct water vapor feature at 1.4 $\mu$m (driving a mixing ratio upper limit of $10^{-5}$) and a high CO abundance ($\sim$20\% of the total atmospheric composition) driven by the excess emission within the Spitzer IRAC 4.5 and 5.8 $\mu$m photometric channels.  

In contrast, \cite{Arcangeli2018} used a 1D-RCTE grid retrieval based analysis on the same data and found a near solar composition metallicity and an upper limit on the C/O $\sim$ 0.8.  They argue that the lack of water features over the HST WFC3 band pass is a natural consequence of the of thermal dissociation of water and onset of H- bound-free continuum opacity--both expected at WASP-18\,b dayside temperatures.  

Our results, in a way, are somewhat consistent with both \cite{Sheppard2017} and \cite{Arcangeli2018}.  If we just consider the CO constraints within the free retrieval framework (similar to \cite{Sheppard2017}), high C/O, high metallicity solutions are possible. However, owing to the higher resolution of IGRINS and coverage of an additional, stronger, water vapor band (see Figure \ref{fig:spectra_components}), we are able to place a constraint on the water mixing ratio--at values generally larger than the \cite{Sheppard2017} upper limit--but also we are sensitive to OH.  These bounded ``oxygen-only" species constraints ``dilute" the C from CO to produce C/O $<$ 1. Considering the partitioning of 40\% of the oxygen into O, this reduces the C/O even further--to solar values.

Our grid based retrieval results are most similar to the \cite{Arcangeli2018} results. The retrieved metallicity is about a half dex higher than their retrieved median, but still falls within about 1.5$\sigma$ of our solution.  Their C/O upper limit contains our grid retrieval solution. Of notable difference is the higher redistribution factor we find compared to \cite{Arcangeli2018}: they find values between $\sim$2 and 2.2, fairly well bounded, whereas the IGRINS data prefer values closer to the max redistribution upper limit (prior edge) of 2.67.  However, as discussed above, the latter seems to have little influence on the retrieved elemental abundances.  



\section{Conclusions}\label{sec:conclusions}
In this work, we present the analysis of ground-based, high resolution emission spectra ($R$=$45,000$) of WASP-18\,b, obtained with IGRINS at Gemini-South and a relatively modest investment of telescope time (2.85 hrs). This represents the first result from our ``Roasting Marshmallows" hot Jupiter survey with IGRINS. We apply state-of-the-art HRCCS analysis techniques and learn the following about the planet's thermal and chemical properties:
\begin{itemize}
    \item We measure three molecular species (H$_2$O, CO, OH) in cross correlation (\S\ref{sec:cc_maps}) and with the likelihood framework (\S\ref{sec:lh_maps}). These are also the only three species with bounded constraints from the free retrieval (\S\ref{sec:retrieval}).
    \item We unambiguously confirm the presence of a thermal inversion layer, via the positive correlation with 1D-RCTE models containing emission lines (\S\ref{sec:cc_maps}) and via the retrieval of a strictly-positive log($\gamma_V$) parameter for the planet's TP profile (\S\ref{sec:retrieval}).
    \item In contrast to emission spectroscopy of the hot Jupiter WASP-77\,Ab \citep{line2021}, we observe a strong degeneracy between a global scaling factor and the parameters describing the TP profile. Such degeneracy prevents us from obtaining meaningful constraints on the planet's irradiation temperature (\S\ref{sec:retrieval}). It is still unclear whether the lifting of the degeneracy in previous work was due to the much stronger detection, or due to the different irradiation regime of the planet.
    \item In spite of the above degeneracy, retrieved absolute abundances appear robust against the value of the scaling factor across orders of magnitude.
    \item We highlight the shortcomings of a free-chemistry model assuming constant vertical abundances and ignoring atomic oxygen to reproduce both the correct line shape and depth in the presence of thermal dissociation (\S\ref{sec:disc_chemistry}). Such model points to a C-to-O ratio (C/O) of $0.75^{+0.14}_{-0.17}$ and a metallicity [M/H]$ = 1.03^{+0.66}_{-1.01}$ for WASP-18\,b, both higher than those of the parent star (C/O=0.23, [M/H]=0.05). 
    \item When we account for $\approx40$\% unmeasured oxygen due to the thermal dissociation of H$_2$O and OH, we obtain a lower C/O$=0.45^{+0.08}_{-0.10}$ but a similar [M/H]$=1.17^{+0.66}_{-1.01}$.
    \item Retrieving C/O and [M/H] with a self-consistent model incorporating the TP-abundance profiles in the presence of thermal dissociation leads to C/O$<$0.34 (2$\sigma$) and [M/H]=0.48$^{+0.33}_{-0.29}$, i.e. tighter constraints in line with stellar values (\S\ref{sec:grid_retrieval}). 
    \item The resulting 1D-RCTE grid retrieval volume mixing ratio and temperature pressure profiles are in agreement with the free retrieval 95\% confidence intervals (\S\ref{sec:grid_retrieval}, Figure \ref{fig:AbundHist_Grid}). The flexibility of the free retrieval allows a broad range of abundance combinations, some nonphysical, that drive the C/O to higher values. 
    \item We see a tentative evidence for additional Doppler shifts compared to the planet's orbital velocity as a function of molecular species, and we advocate for follow-up observations to confirm these shifts (\S\ref{sec:disc_vel_shift}).
\end{itemize}

As we develop the systematic application of HRCCS to robustly retrieve abundances and temperature of exoplanets, using infrared spectroscopy of an ultra-hot Jupiter we highlight a few final points relevant to future work.
Firstly, given the strong degeneracy with the TP profile, the inclusion of a log($a$) parameter appears conceptually questionable. In fact, an appropriate atmospheric model should reproduce the correct line depth by adjusting the other physical parameters. On the other hand, directly fitting for log($a$) allows us to diagnose potential modelling shortcomings (as in this case) when log($a$) is significantly deviant from 0. 
Secondly, UHJs require a more complex parametrization of the abundance profiles of (at least) H$_2$O and OH, and an estimate of atomic oxygen produced via thermal dissociation \citep[see also][]{Kasper2022}. This is to be implemented in future work.
Lastly, putting together the lessons learned from ground and space observations of WASP-18\,b, we anticipate that the interpretation of JWST emission spectra of UHJs will be complex, especially if CO is weakly detectable and OH and O are not detectable. On the other hand, these spectra will not suffer from the TP profile degeneracy due to the additional continuum information retained. 

Overall, the combination of low-resolution and high-resolution spectroscopy is particularly appealing for this class of planets, particularly for WASP-18\,b, as it is a JWST Early Release Science target and is being observed at the time of writing.

\begin{acknowledgements}
This work used the Immersion Grating Infrared Spectrometer (IGRINS) that was developed under a collaboration between the University of Texas at Austin and the Korea Astronomy and Space Science Institute (KASI) with the financial support of the Mt. Cuba Astronomical Foundation, of the US National Science Foundation under grants AST-1229522 and AST-1702267, of the McDonald Observatory of the University of Texas at Austin, of the Korean GMT Project of KASI, and Gemini Observatory. J.L.Bean\, M.R.L.\, and P.S.\ acknowledge support from NASA XRP grant 80NSSC19K0293. M.B.\ acknowledges support from the STFC research grant ST/T000406/1. M.M. acknowledges support through NASA Hubble Fellowship grant HST-HF2-51485.001-A awarded by the Space Telescope Science Institute, which is operated by the Association of Universities for Research in Astronomy, Inc., for NASA, under contract NAS5-26555. J. L. Birkby acknowledges funding from the European Research Council (ERC) under the European Union’s Horizon 2020 research and innovation program under grant agreement No 805445.  We would also like to thank the NOIRLabs support staff helping with the implementation of these observations.  Finally, we acknowledge Research Computing at Arizona State University for providing HPC and storage resources that have significantly contributed to the research results reported within this manuscript.
\end{acknowledgements}

\facilities{Gemini-South (IGRINS)}
\software{\texttt{astropy} \citep{astropy:2018}, \texttt{barycorrpy} \citep{barycorrpy}, \texttt{corner} \citep{corner}, \texttt{matplotlib} \citep{matplotlib:2007}, \texttt{numpy} \citep{numpy:2020}, \texttt{python} \citep{python3:2009}, \texttt{scipy} \citep{2020SciPy-NMeth}}

\newpage

\appendix
\section{Robustness of the detection to a range of telluric removal algorithms}\label{sec:pca_vs_svd}

To our knowledge, a full quantitative comparison of telluric removal algorithms has never been performed before. In \citet{line2021}, we have investigated the dependence on the number of SVD components removed, but in this study we address additional subtleties that can potentially affect the result.

\begin{figure*}[h]
\begin{center}
\includegraphics[width=0.45\textwidth]{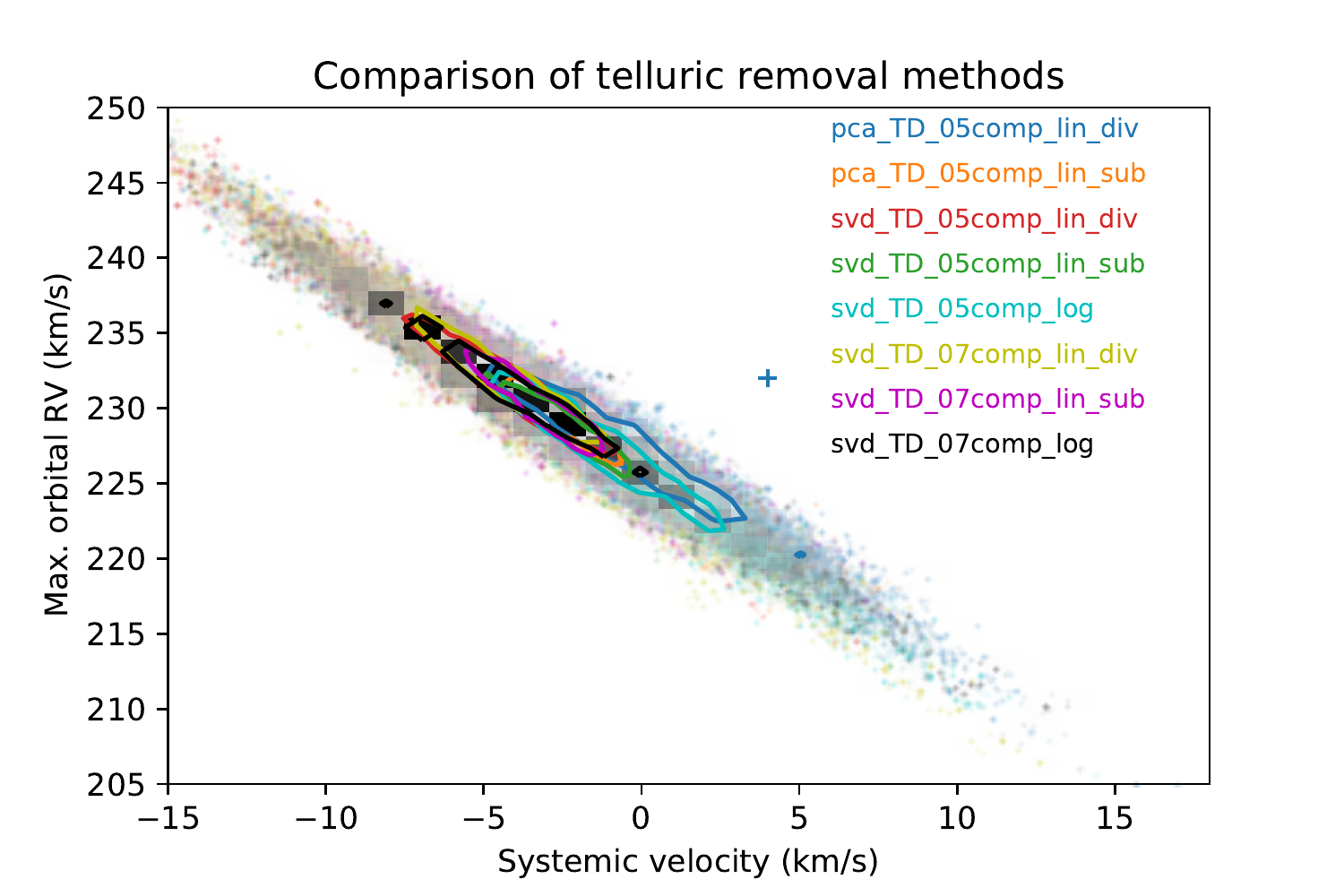}
\includegraphics[width=0.45\textwidth]{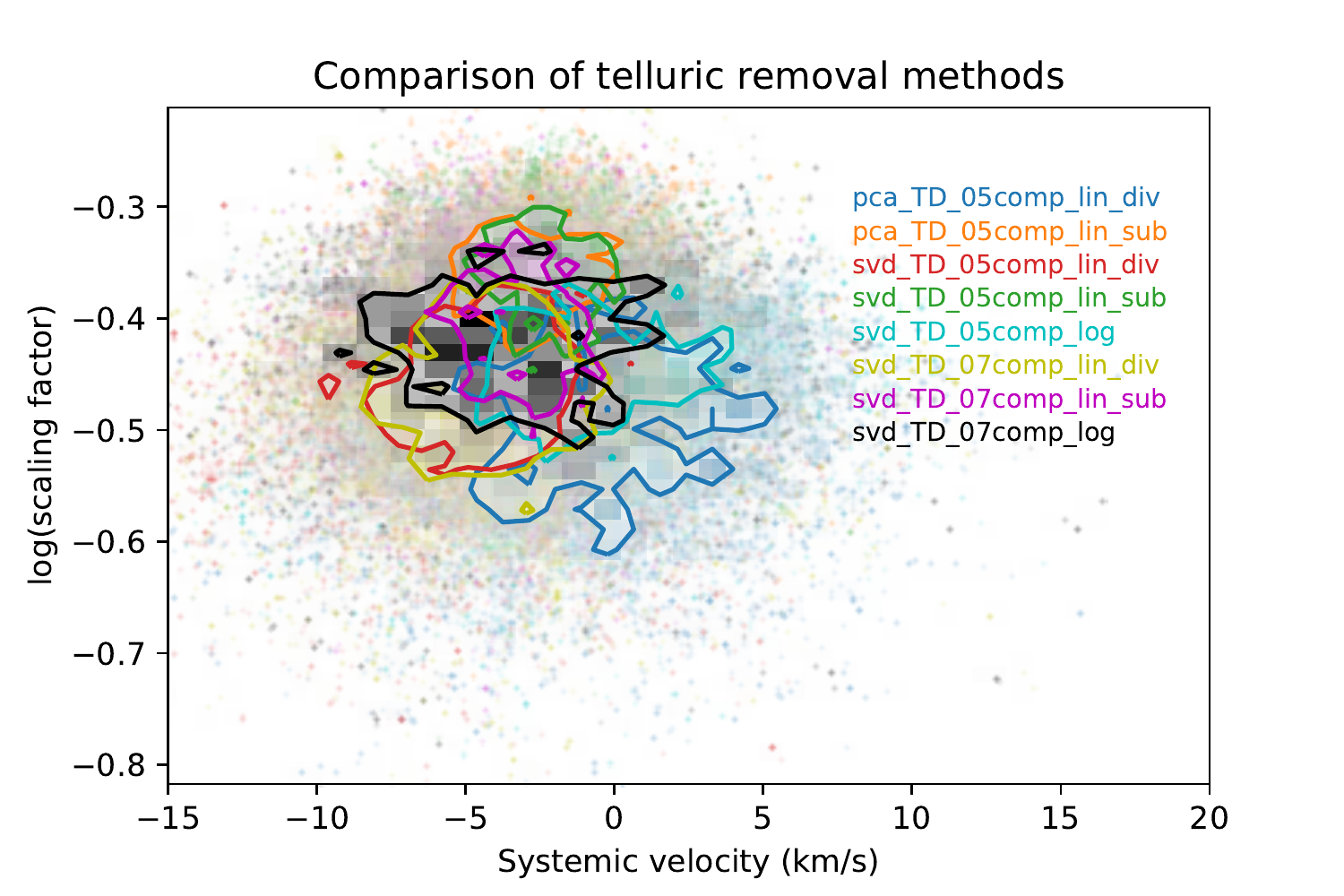}
\caption{Posteriors in $V_\mathrm{sys}, K_\mathrm{P}$ (left panel) or in $V_\mathrm{sys}, \log(a)$ (right panel) obtained by varying the telluric removal analysis. Labels encode type of analysis done (\texttt{pca} or \texttt{svd}) in the time domain (\texttt{TD}), the number of components (\texttt{XXcomp}), the flux space (\texttt{lin} for linear, \texttt{log} for logarithmic), and whether the fit is divided (\texttt{div}) through the data or subtracted (\texttt{sub}) out. It shows that the best fit parameters are unbiased within 1$\sigma$ (solid contours) regardless of the subtleties of the telluric removal.}
\label{fig:robustness_analysis}
\end{center}
\end{figure*}

To make the comparison quantitative, we select the equilibrium model containing the three detected species (H$_2$O, CO, OH; see Section~\ref{sec:CCF_analysis}) and we explore the 3 parameter space in ($V_\mathrm{sys}, K_\mathrm{P}, \log(a)$) as in \S\ref{sec:lh_maps}, for each of the tested analysis. The posteriors of the three parameters are then compared quantitatively to look for significant shifts in their best-fit values. Figure~\ref{fig:robustness_analysis} presents a representative subset of the 120 tests we performed. We compares the SVD-only algorithm of \citet{line2021} with the full Principal Component Analysis (PCA) of \citet{Giacobbe2021}, which contains SVD as one of the three main steps. We ran the algorithms in both the time domain (TD) and wavelength domain (WD), varying the number of components between 3 and 8. Furthermore, we ran the telluric removal in both logarithmic and linear flux space. In the latter case, we compare subtracting the best-fit to dividing through the best fit. We note that changing the flux space and correction method is equivalent to weighting the data by their relative vs absolute error. 

The signal from the planet is detected in all cases, albeit the size of the confidence intervals varies, indicating that some analyses seem to be more effective at preserving the exoplanet signal than other. Furthermore, within 1$\sigma$ all the versions of the analysis yield the same best-fit parameters, as shown in Figure~\ref{fig:robustness_analysis}. We note that this result relies on model reprocessing to be applied and modified accordingly to the type of the analysis selected, so that model and data undergo the same processing.

Incidentally, the smaller confidence intervals are obtained with the analysis presented in this study (SVD, time domain, linear space, subtraction), which is also the analysis used in \citet{line2021}.

\section{HyDRA-H and CHIMERA Retrieval Comparison}

We bench-marked the retrieval results of CHIMERA against a retrieval performed on WASP-18\~b with HyDRA-H \citep{Gandhi19}, similarly to what we did for WASP-77\,Ab in \citet{line2021}. The HyDRA-H retrieval used 13 free parameters, namely the volume mixing ratios of the chemical species H$_2$O, OH, CO, H- and Ca, 6 free parameters to describe the temperature profile using the prescription of \citet{Madhusudhan2009}, and 2 parameters for velocity shifts from the expected values of K$_\mathrm{P}$ and V$_\mathrm{rest}$. For the molecular and atomic cross sections we used the HITEMP database \citep{hitemp2010} for H$_2$O, CO and OH, the Kurucz database for Ca \citep{kurucz1995} and \citet{bell1987} and \citet{john1988} for H- see also \citep[see also][]{Gandhi2020}. Our statistical analysis was performed with the MultiNest Nested Sampling algorithm \citep{feroz2008, feroz2009, Buchner2014}. Further details of the retrieval setup can be found in \citet{Gandhi19} and \citet{line2021}. This retrieval can be directly compared with the ``Fixed $a$=1'' retrieval presented in \S\ref{sec:testing_loga} and summarized in Table~\ref{tab:posteriors}. We note that the parametrization of the TP profile is different between the two retrievals, and therefore this is a good test to highlight the dependence of the result on the parametrization for UHJs, which might differ from the agreement measured for a hot Jupiter such as WASP-77\,Ab.

\begin{figure*}[h]
\begin{center}
\includegraphics[width=0.45\textwidth]{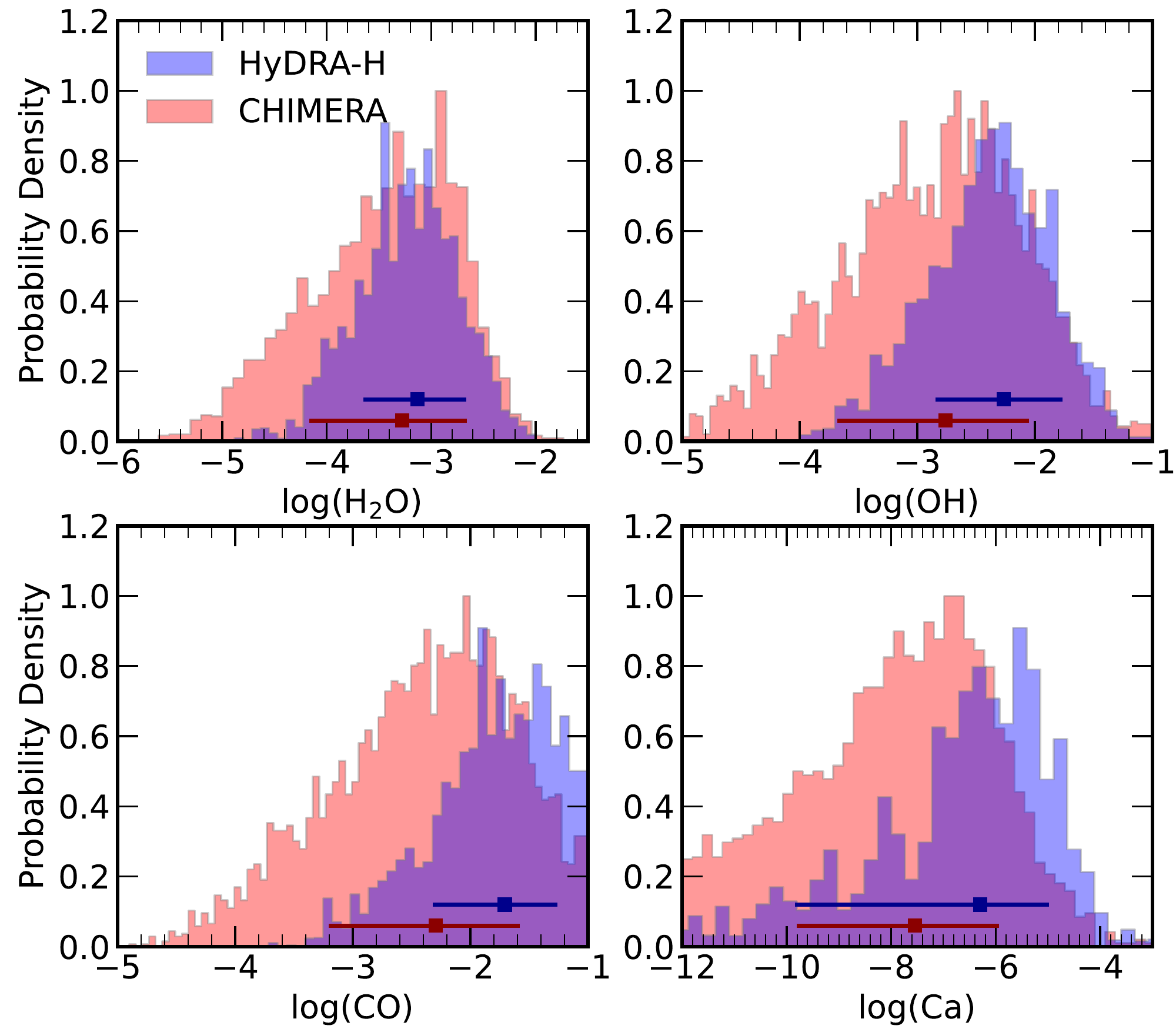}
\includegraphics[width=0.45\textwidth]{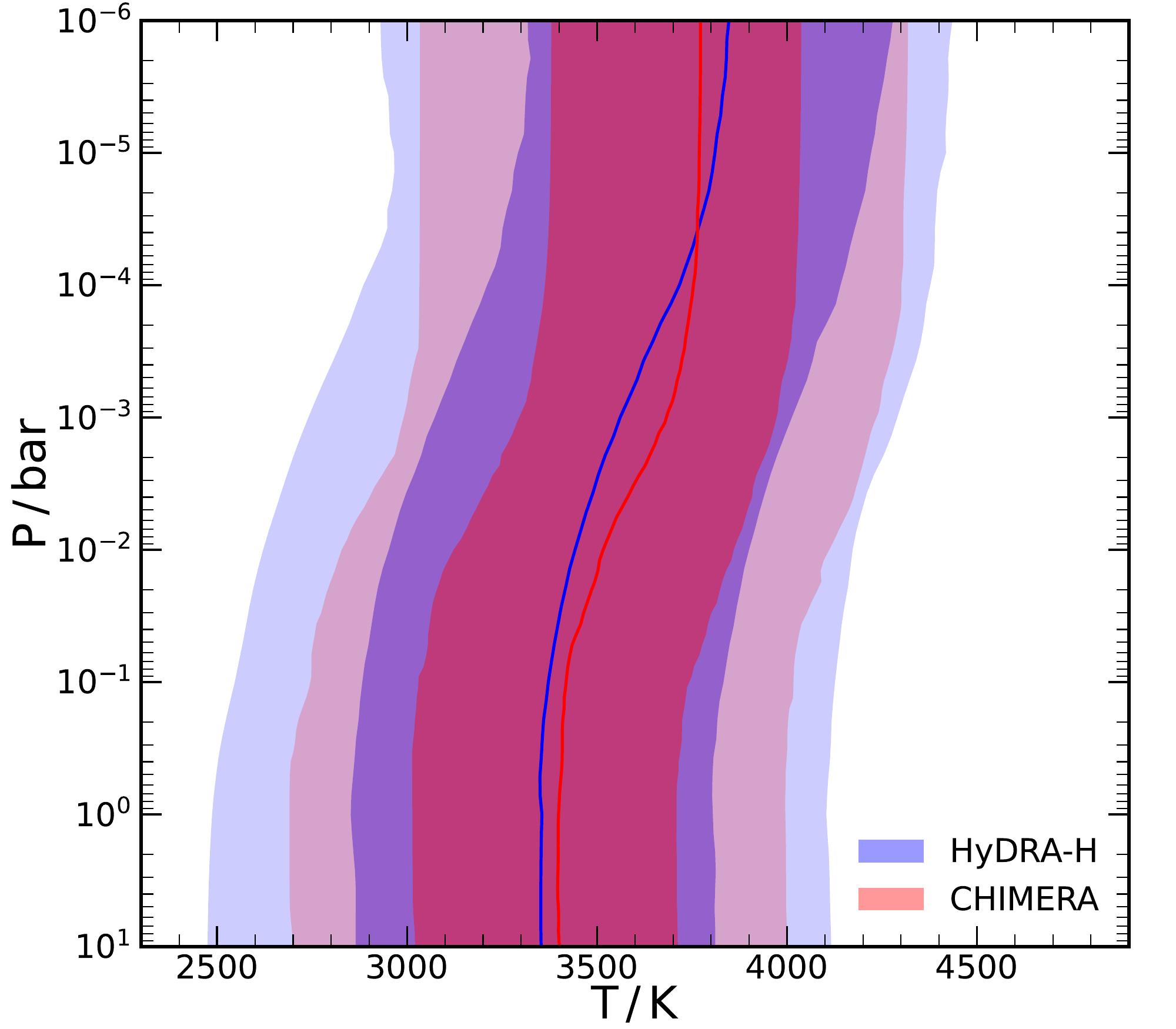}
\caption{Comparison between the HyDRA-H retrieval (in blue) and the ``Fixed $a$=1'' CHIMERA retrieval (in red, see also Table~\ref{tab:posteriors} and \S\ref{sec:testing_loga}). Constraints on abundances (four panels, left side) and the TP profile (right side) are compatible within 1$\sigma$.}
\label{fig:compare}
\end{center}
\end{figure*}

The retrieved volume mixing ratios are shown in Figure~\ref{fig:compare} and show good agreement with all of the constraints from CHIMERA. We strongly constrain H$_2$O, OH and CO, but only obtain an upper limit for Ca due to the weaker peak in the posterior. Our temperature profiles are also in good agreement, indicating temperatures near 3500~K for the photosphere with an inversion of $\sim500$~K. However, there is a very slight difference in the temperature gradient, the HyDRA-H retrievals constrain a shallower inversion which requires a slightly higher abundance of the chemical species to fit the spectrum. However, these differences in the temperature structure and abundances is well within the 1$\sigma$ error bars of each of the retrievals, and the overall agreement between the two different models remains excellent.

\section{Corner plots from the 5 retrievals of WASP-18\,b}\label{sec:app_corner_plots}

In this section we provide the full corner plots obtained from the 5 retrieval runs described in \S\ref{sec:retrieval}, as well as the full corner plot for the grid retrieval described in \S\ref{sec:grid_retrieval}.

\begin{figure*}
\begin{center}
\includegraphics[width=\textwidth]{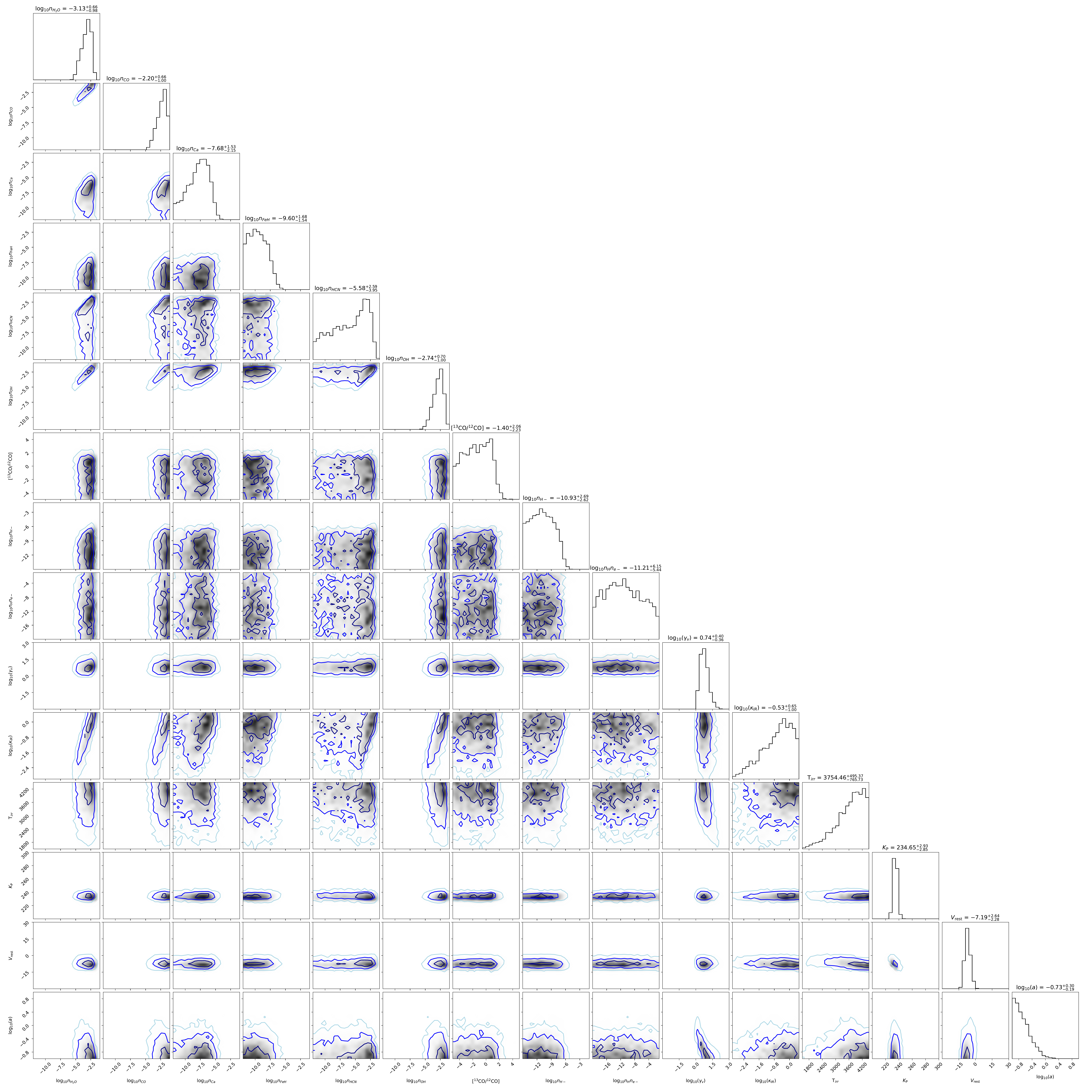}
\caption{Corner plots for the 15 parameters of the ``fiducial'' retrieval including a scaling factor. See \S\ref{sec:retrieval} for details.}
\label{fig:corner_full_15pars}
\end{center}
\end{figure*}

\begin{figure*}
\begin{center}
\includegraphics[width=\textwidth]{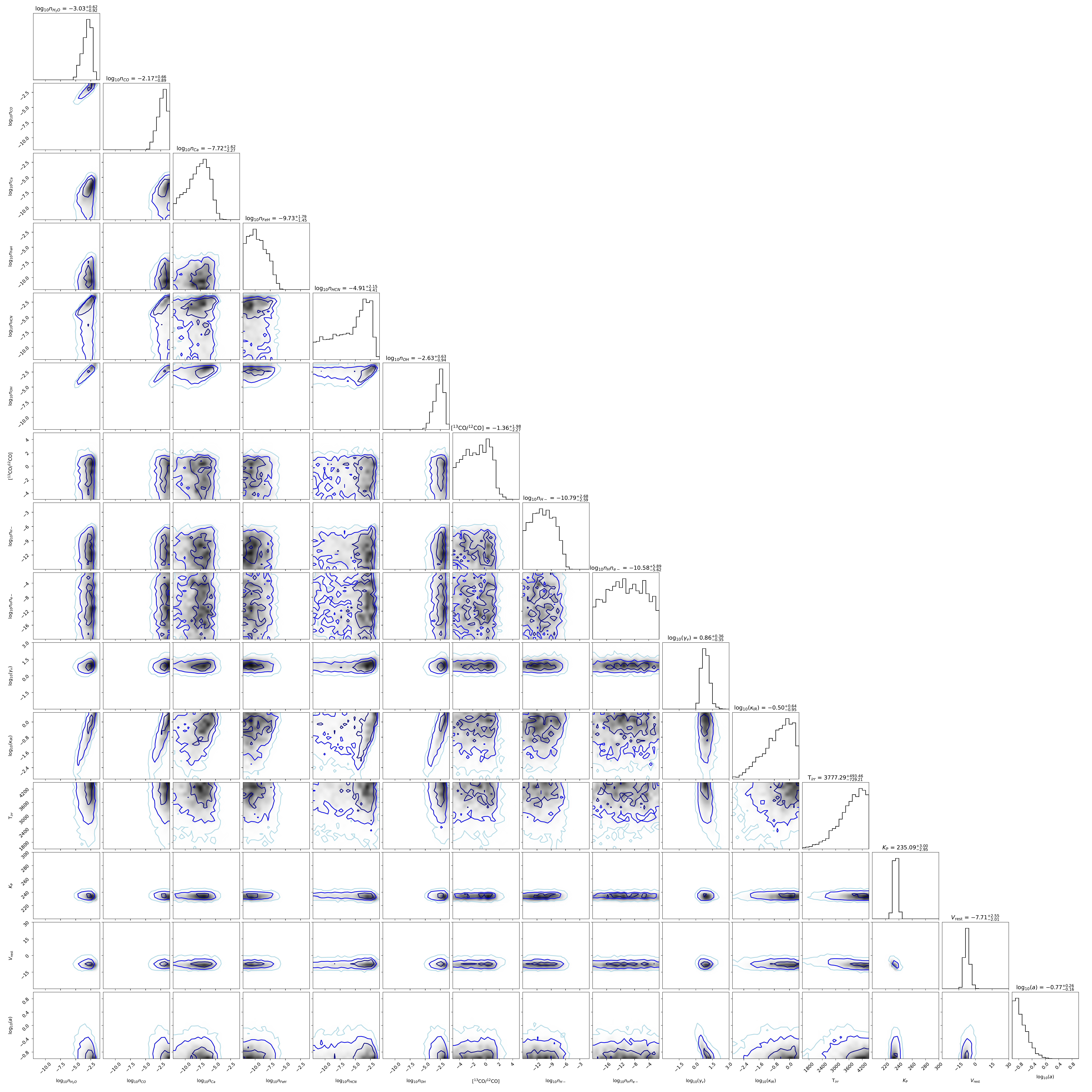}
\caption{Corner plots for the 15 parameters of the retrieval where a cosine function is applied to the scaling factor as a function of planet orbital phase (labelled as ``Phase Dep. $a$'' in Table \ref{tab:posteriors}). See \S\ref{sec:testing_loga} for details.}
\label{fig:corner_phase-dep}
\end{center}
\end{figure*}

\begin{figure*}
\begin{center}
\includegraphics[width=\textwidth]{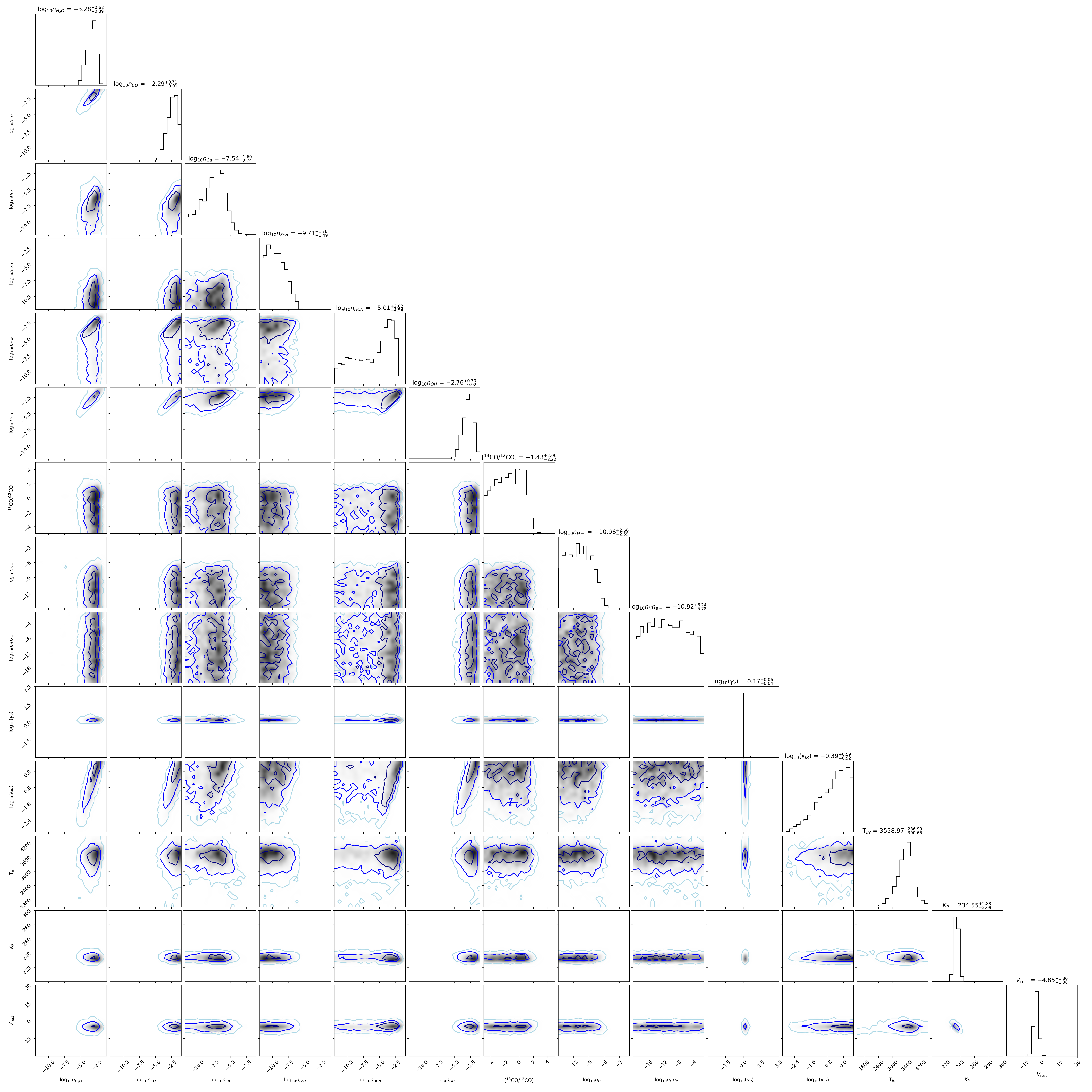}
\caption{Corner plots for the 14 parameters of the retrieval with the scaling factor set to 1 (labelled as ``Fixed $a$=1'' in Table \ref{tab:posteriors}). See \S\ref{sec:testing_loga} for details.}
\label{fig:corner_full_14pars}
\end{center}
\end{figure*}

\begin{figure*}
\begin{center}
\includegraphics[width=\textwidth]{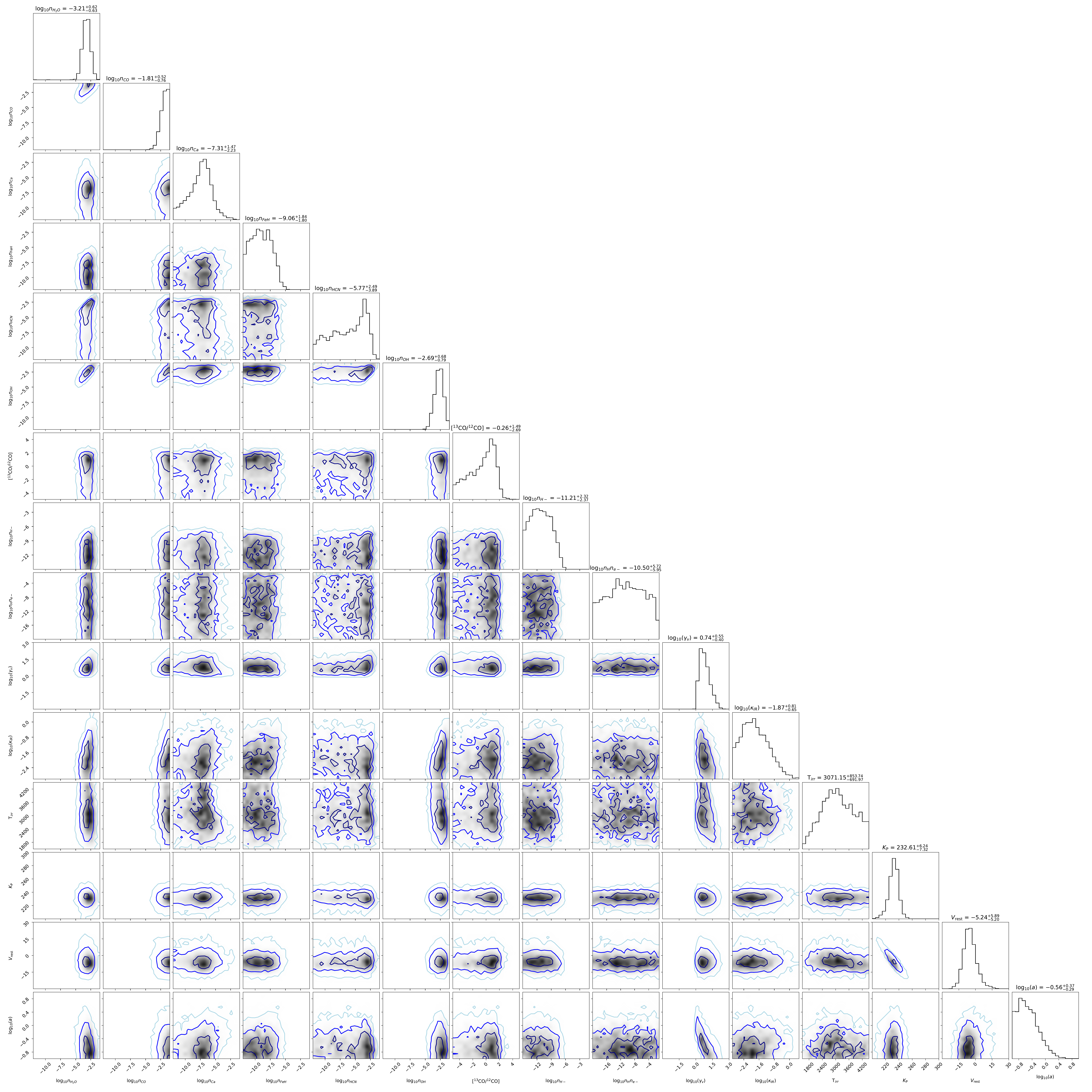}
\caption{Corner plots for the 15 parameters of the retrieval including only the first half of the spectra (labelled as ``1$^\mathrm{st}$ half'' in Table \ref{tab:posteriors}). See \S\ref{sec:phase_resolved} for details.}
\label{fig:corner_1st_half}
\end{center}
\end{figure*}

\begin{figure*}
\begin{center}
\includegraphics[width=\textwidth]{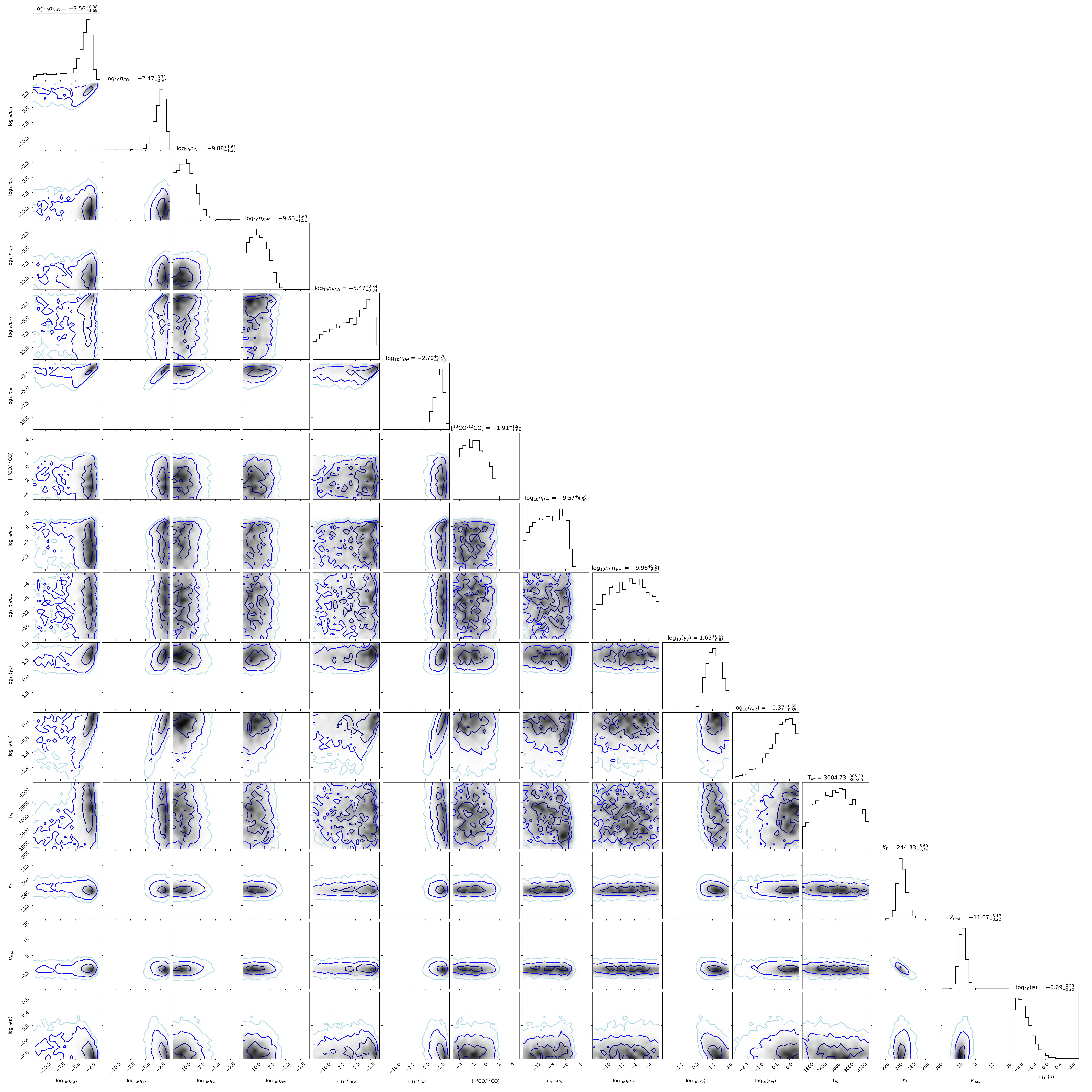}
\caption{Corner plots for the 15 parameters of the retrieval including only the second half of the spectra (labelled as ``2$^\mathrm{nd}$ half'' in Table \ref{tab:posteriors}). See \S\ref{sec:phase_resolved} for details.}
\label{fig:corner_2nd_half}
\end{center}
\end{figure*}

\begin{figure*}
\begin{center}
\includegraphics[width=\textwidth]{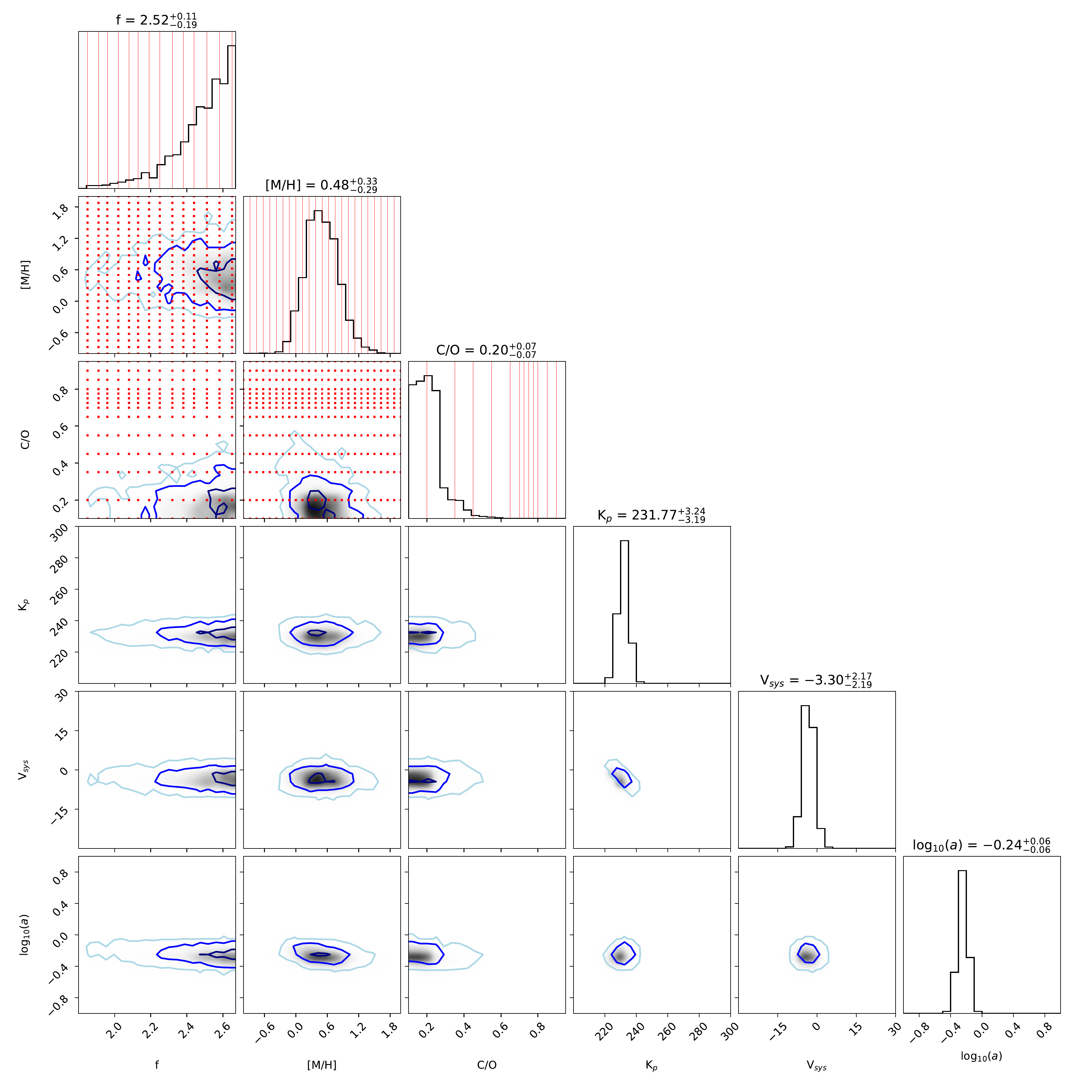}
\caption{Corner plot for the 6 parameters of the grid retrieval including a scaling factor. The red dots mark the grid points corresponding to each of the computed 1D RCTE models. The same grid points are marked for each parameter as solid red lines in the marginalised posteriors. See \S\ref{sec:grid_retrieval} for details.}
\label{fig:corner_grid_retrieval}
\end{center}
\end{figure*}

\bibliography{references}{}

\begin{thebibliography}{}
\expandafter\ifx\csname natexlab\endcsname\relax\def\natexlab#1{#1}\fi
\providecommand{\url}[1]{\href{#1}{#1}}
\providecommand{\dodoi}[1]{doi:~\href{http://doi.org/#1}{\nolinkurl{#1}}}
\providecommand{\doeprint}[1]{\href{http://ascl.net/#1}{\nolinkurl{http://ascl.net/#1}}}
\providecommand{\doarXiv}[1]{\href{https://arxiv.org/abs/#1}{\nolinkurl{https://arxiv.org/abs/#1}}}

\bibitem[{{Arcangeli} {et~al.}(2018){Arcangeli}, {D{\'e}sert}, {Line}, {Bean},
  {Parmentier}, {Stevenson}, {Kreidberg}, {Fortney}, {Mansfield}, \&
  {Showman}}]{Arcangeli2018}
{Arcangeli}, J., {D{\'e}sert}, J.-M., {Line}, M.~R., {et~al.} 2018, \apjl, 855,
  L30, \dodoi{10.3847/2041-8213/aab272}

\bibitem[{{Arcangeli} {et~al.}(2019){Arcangeli}, {D{\'e}sert}, {Parmentier},
  {Stevenson}, {Bean}, {Line}, {Kreidberg}, {Fortney}, \&
  {Showman}}]{Arcangeli2019}
{Arcangeli}, J., {D{\'e}sert}, J.-M., {Parmentier}, V., {et~al.} 2019, \aap,
  625, A136, \dodoi{10.1051/0004-6361/201834891}

\bibitem[{{Astropy Collaboration} {et~al.}(2018){Astropy Collaboration},
  {Price-Whelan}, {Sip{\H{o}}cz}, {G{\"u}nther}, {Lim}, {Crawford}, {Conseil},
  {Shupe}, {Craig}, {Dencheva}, {Ginsburg}, {Vand erPlas}, {Bradley},
  {P{\'e}rez-Su{\'a}rez}, {de Val-Borro}, {Aldcroft}, {Cruz}, {Robitaille},
  {Tollerud}, {Ardelean}, {Babej}, {Bach}, {Bachetti}, {Bakanov}, {Bamford},
  {Barentsen}, {Barmby}, {Baumbach}, {Berry}, {Biscani}, {Boquien}, {Bostroem},
  {Bouma}, {Brammer}, {Bray}, {Breytenbach}, {Buddelmeijer}, {Burke},
  {Calderone}, {Cano Rodr{\'\i}guez}, {Cara}, {Cardoso}, {Cheedella}, {Copin},
  {Corrales}, {Crichton}, {D'Avella}, {Deil}, {Depagne}, {Dietrich}, {Donath},
  {Droettboom}, {Earl}, {Erben}, {Fabbro}, {Ferreira}, {Finethy}, {Fox},
  {Garrison}, {Gibbons}, {Goldstein}, {Gommers}, {Greco}, {Greenfield},
  {Groener}, {Grollier}, {Hagen}, {Hirst}, {Homeier}, {Horton}, {Hosseinzadeh},
  {Hu}, {Hunkeler}, {Ivezi{\'c}}, {Jain}, {Jenness}, {Kanarek}, {Kendrew},
  {Kern}, {Kerzendorf}, {Khvalko}, {King}, {Kirkby}, {Kulkarni}, {Kumar},
  {Lee}, {Lenz}, {Littlefair}, {Ma}, {Macleod}, {Mastropietro}, {McCully},
  {Montagnac}, {Morris}, {Mueller}, {Mumford}, {Muna}, {Murphy}, {Nelson},
  {Nguyen}, {Ninan}, {N{\"o}the}, {Ogaz}, {Oh}, {Parejko}, {Parley}, {Pascual},
  {Patil}, {Patil}, {Plunkett}, {Prochaska}, {Rastogi}, {Reddy Janga},
  {Sabater}, {Sakurikar}, {Seifert}, {Sherbert}, {Sherwood-Taylor}, {Shih},
  {Sick}, {Silbiger}, {Singanamalla}, {Singer}, {Sladen}, {Sooley},
  {Sornarajah}, {Streicher}, {Teuben}, {Thomas}, {Tremblay}, {Turner},
  {Terr{\'o}n}, {van Kerkwijk}, {de la Vega}, {Watkins}, {Weaver}, {Whitmore},
  {Woillez}, {Zabalza}, \& {Astropy Contributors}}]{astropy:2018}
{Astropy Collaboration}, {Price-Whelan}, A.~M., {Sip{\H{o}}cz}, B.~M., {et~al.}
  2018, \aj, 156, 123, \dodoi{10.3847/1538-3881/aabc4f}

\bibitem[{Barber {et~al.}(2013)Barber, Strange, Hill, Polyansky, Mellau,
  Yurchenko, \& Tennyson}]{BarberHCN}
Barber, R.~J., Strange, J.~K., Hill, C., {et~al.} 2013, Monthly Notices of the
  Royal Astronomical Society, 437, 1828, \dodoi{10.1093/mnras/stt2011}

\bibitem[{{Baxter} {et~al.}(2020){Baxter}, {D{\'e}sert}, {Parmentier}, {Line},
  {Fortney}, {Arcangeli}, {Bean}, {Todorov}, \& {Mansfield}}]{Baxter2020}
{Baxter}, C., {D{\'e}sert}, J.-M., {Parmentier}, V., {et~al.} 2020, \aap, 639,
  A36, \dodoi{10.1051/0004-6361/201937394}

\bibitem[{{Bean} {et~al.}(2018){Bean}, {Stevenson}, {Batalha},
  {Berta-Thompson}, {Kreidberg}, {Crouzet}, {Benneke}, {Line}, {Sing}, \&
  {Wakeford}}]{Bean2018}
{Bean}, J.~L., {Stevenson}, K.~B., {Batalha}, N.~M., {et~al.} 2018, \pasp, 130,
  114402, \dodoi{10.1088/1538-3873/aadbf3}

\bibitem[{{Bedell} {et~al.}(2018){Bedell}, {Bean}, {Mel{\'e}ndez}, {Spina},
  {Ram{\'\i}rez}, {Asplund}, {Alves-Brito}, {dos Santos}, {Dreizler}, {Yong},
  {Monroe}, \& {Casagrande}}]{Bedell2018}
{Bedell}, M., {Bean}, J.~L., {Mel{\'e}ndez}, J., {et~al.} 2018, \apj, 865, 68,
  \dodoi{10.3847/1538-4357/aad908}

\bibitem[{{Bell} \& {Berrington}(1987)}]{bell1987}
{Bell}, K.~L., \& {Berrington}, K.~A. 1987, Journal of Physics B Atomic
  Molecular Physics, 20, 801, \dodoi{10.1088/0022-3700/20/4/019}

\bibitem[{{Beltz} {et~al.}(2022){Beltz}, {Rauscher}, {Kempton}, {Malsky},
  {Ochs}, {Arora}, \& {Savel}}]{Beltz2022}
{Beltz}, H., {Rauscher}, E., {Kempton}, E. M.~R., {et~al.} 2022, \aj, 164, 140,
  \dodoi{10.3847/1538-3881/ac897b}

\bibitem[{Bernath(2020)}]{Bernath2020MOLLIST}
Bernath, P.~F. 2020, \jqsrt, 240, 106687,
  \dodoi{https://doi.org/10.1016/j.jqsrt.2019.106687}

\bibitem[{{Birkby}(2018)}]{Birkby2018}
{Birkby}, J.~L. 2018, arXiv e-prints, arXiv:1806.04617.
\newblock \doarXiv{1806.04617}

\bibitem[{{Brewer} \& {Fischer}(2016)}]{Brewer16}
{Brewer}, J.~M., \& {Fischer}, D.~A. 2016, \apj, 831, 20,
  \dodoi{10.3847/0004-637X/831/1/20}

\bibitem[{{Brogi} \& {Birkby}(2021)}]{BrogiBirkby2021}
{Brogi}, M., \& {Birkby}, J. 2021, in ExoFrontiers; Big Questions in
  Exoplanetary Science, ed. N.~{Madhusudhan}, 8--1,
  \dodoi{10.1088/2514-3433/abfa8fch8}

\bibitem[{{Brogi} \& {Line}(2019)}]{BL19}
{Brogi}, M., \& {Line}, M.~R. 2019, \aj, 157, 114,
  \dodoi{10.3847/1538-3881/aaffd3}

\bibitem[{Brogi \& Line(2019)}]{Brogi2019}
Brogi, M., \& Line, M.~R. 2019, AJ, 157, 114

\bibitem[{{Buchner} {et~al.}(2014){Buchner}, {Georgakakis}, {Nandra}, {Hsu},
  {Rangel}, {Brightman}, {Merloni}, {Salvato}, {Donley}, \&
  {Kocevski}}]{Buchner2014}
{Buchner}, J., {Georgakakis}, A., {Nandra}, K., {et~al.} 2014, \aap, 564, A125,
  \dodoi{10.1051/0004-6361/201322971}

\bibitem[{{Burrows} {et~al.}(2006){Burrows}, {Sudarsky}, \&
  {Hubeny}}]{Burrows2006}
{Burrows}, A., {Sudarsky}, D., \& {Hubeny}, I. 2006, \apj, 650, 1140,
  \dodoi{10.1086/507269}

\bibitem[{{Casasayas-Barris} {et~al.}(2019){Casasayas-Barris}, {Pall{\'e}},
  {Yan}, {Chen}, {Kohl}, {Stangret}, {Parviainen}, {Helling}, {Watanabe},
  {Czesla}, {Fukui}, {Monta{\~n}{\'e}s-Rodr{\'\i}guez}, {Nagel}, {Narita},
  {Nortmann}, {Nowak}, {Schmitt}, \& {Zapatero Osorio}}]{Casasayas19}
{Casasayas-Barris}, N., {Pall{\'e}}, E., {Yan}, F., {et~al.} 2019, \aap, 628,
  A9, \dodoi{10.1051/0004-6361/201935623}

\bibitem[{{Cont} {et~al.}(2021){Cont}, {Yan}, {Reiners}, {Casasayas-Barris},
  {Molli{\`e}re}, {Pall{\'e}}, {Henning}, {Nortmann}, {Stangret}, {Czesla},
  {L{\'o}pez-Puertas}, {S{\'a}nchez-L{\'o}pez}, {Rodler}, {Ribas},
  {Quirrenbach}, {Caballero}, {Amado}, {Carone}, {Khaimova}, {Kreidberg},
  {Molaverdikhani}, {Montes}, {Morello}, {Nagel}, {Oshagh}, \&
  {Zechmeister}}]{cont21}
{Cont}, D., {Yan}, F., {Reiners}, A., {et~al.} 2021, \aap, 651, A33,
  \dodoi{10.1051/0004-6361/202140732}

\bibitem[{{Cowan} \& {Agol}(2011)}]{Cowan2011}
{Cowan}, N.~B., \& {Agol}. 2011, \apj, 729, 54,
  \dodoi{10.1088/0004-637X/729/1/54}

\bibitem[{{de Kok} {et~al.}(2013){de Kok}, {Brogi}, {Snellen}, {Birkby},
  {Albrecht}, \& {de Mooij}}]{DeKok2013}
{de Kok}, R.~J., {Brogi}, M., {Snellen}, I.~A.~G., {et~al.} 2013, \aap, 554,
  A82, \dodoi{10.1051/0004-6361/201321381}

\bibitem[{{Ehrenreich} {et~al.}(2020){Ehrenreich}, {Lovis}, {Allart}, {Zapatero
  Osorio}, {Pepe}, {Cristiani}, {Rebolo}, {Santos}, {Borsa}, {Demangeon},
  {Dumusque}, {Gonz{\'a}lez Hern{\'a}ndez}, {Casasayas-Barris},
  {S{\'e}gransan}, {Sousa}, {Abreu}, {Adibekyan}, {Affolter}, {Allende Prieto},
  {Alibert}, {Aliverti}, {Alves}, {Amate}, {Avila}, {Baldini}, {Bandy}, {Benz},
  {Bianco}, {Bolmont}, {Bouchy}, {Bourrier}, {Broeg}, {Cabral}, {Calderone},
  {Pall{\'e}}, {Cegla}, {Cirami}, {Coelho}, {Conconi}, {Coretti}, {Cumani},
  {Cupani}, {Dekker}, {Delabre}, {Deiries}, {D'Odorico}, {Di Marcantonio},
  {Figueira}, {Fragoso}, {Genolet}, {Genoni}, {G{\'e}nova Santos}, {Hara},
  {Hughes}, {Iwert}, {Kerber}, {Knudstrup}, {Land oni}, {Lavie}, {Lizon},
  {Lendl}, {Lo Curto}, {Maire}, {Manescau}, {Martins}, {M{\'e}gevand },
  {Mehner}, {Micela}, {Modigliani}, {Molaro}, {Monteiro}, {Monteiro},
  {Moschetti}, {M{\"u}ller}, {Nunes}, {Oggioni}, {Oliveira}, {Pariani},
  {Pasquini}, {Poretti}, {Rasilla}, {Redaelli}, {Riva}, {Santana Tschudi},
  {Santin}, {Santos}, {Segovia Milla}, {Seidel}, {Sosnowska}, {Sozzetti},
  {Span{\`o}}, {Su{\'a}rez Mascare{\~n}o}, {Tabernero}, {Tenegi}, {Udry},
  {Zanutta}, \& {Zerbi}}]{Ehrenreich2020}
{Ehrenreich}, D., {Lovis}, C., {Allart}, R., {et~al.} 2020, \nat, 580, 597,
  \dodoi{10.1038/s41586-020-2107-1}

\bibitem[{{Feroz} \& {Hobson}(2008)}]{feroz2008}
{Feroz}, F., \& {Hobson}, M.~P. 2008, \mnras, 384, 449,
  \dodoi{10.1111/j.1365-2966.2007.12353.x}

\bibitem[{{Feroz} {et~al.}(2009){Feroz}, {Hobson}, \& {Bridges}}]{feroz2009}
{Feroz}, F., {Hobson}, M.~P., \& {Bridges}, M. 2009, \mnras, 398, 1601,
  \dodoi{10.1111/j.1365-2966.2009.14548.x}

\bibitem[{Foreman-Mackey(2016)}]{corner}
Foreman-Mackey, D. 2016, The Journal of Open Source Software, 1, 24,
  \dodoi{10.21105/joss.00024}

\bibitem[{{Fortney}(2012)}]{Fortney2012}
{Fortney}, J.~J. 2012, \apjl, 747, L27, \dodoi{10.1088/2041-8205/747/2/L27}

\bibitem[{{Fortney} {et~al.}(2008){Fortney}, {Lodders}, {Marley}, \&
  {Freedman}}]{Fortney2008}
{Fortney}, J.~J., {Lodders}, K., {Marley}, M.~S., \& {Freedman}, R.~S. 2008,
  \apj, 678, 1419, \dodoi{10.1086/528370}

\bibitem[{{Fortney} {et~al.}(2005){Fortney}, {Marley}, {Lodders}, {Saumon}, \&
  {Freedman}}]{Fortney2005a}
{Fortney}, J.~J., {Marley}, M.~S., {Lodders}, K., {Saumon}, D., \& {Freedman},
  R. 2005, \apjl, 627, L69, \dodoi{10.1086/431952}

\bibitem[{{Gaia Collaboration} {et~al.}(2018){Gaia Collaboration}, {Brown},
  {Vallenari}, {Prusti}, {de Bruijne}, {Babusiaux}, {Bailer-Jones}, {Biermann},
  {Evans}, {Eyer}, {Jansen}, {Jordi}, {Klioner}, {Lammers}, {Lindegren},
  {Luri}, {Mignard}, {Panem}, {Pourbaix}, {Randich}, {Sartoretti}, {Siddiqui},
  {Soubiran}, {van Leeuwen}, {Walton}, {Arenou}, {Bastian}, {Cropper},
  {Drimmel}, {Katz}, {Lattanzi}, {Bakker}, {Cacciari}, {Casta{\~n}eda},
  {Chaoul}, {Cheek}, {De Angeli}, {Fabricius}, {Guerra}, {Holl}, {Masana},
  {Messineo}, {Mowlavi}, {Nienartowicz}, {Panuzzo}, {Portell}, {Riello},
  {Seabroke}, {Tanga}, {Th{\'e}venin}, {Gracia-Abril}, {Comoretto},
  {Garcia-Reinaldos}, {Teyssier}, {Altmann}, {Andrae}, {Audard},
  {Bellas-Velidis}, {Benson}, {Berthier}, {Blomme}, {Burgess}, {Busso},
  {Carry}, {Cellino}, {Clementini}, {Clotet}, {Creevey}, {Davidson}, {De
  Ridder}, {Delchambre}, {Dell'Oro}, {Ducourant},
  {Fern{\'a}ndez-Hern{\'a}ndez}, {Fouesneau}, {Fr{\'e}mat}, {Galluccio},
  {Garc{\'\i}a-Torres}, {Gonz{\'a}lez-N{\'u}{\~n}ez}, {Gonz{\'a}lez-Vidal},
  {Gosset}, {Guy}, {Halbwachs}, {Hambly}, {Harrison}, {Hern{\'a}ndez},
  {Hestroffer}, {Hodgkin}, {Hutton}, {Jasniewicz}, {Jean-Antoine-Piccolo},
  {Jordan}, {Korn}, {Krone-Martins}, {Lanzafame}, {Lebzelter}, {L{\"o}ffler},
  {Manteiga}, {Marrese}, {Mart{\'\i}n-Fleitas}, {Moitinho}, {Mora}, {Muinonen},
  {Osinde}, {Pancino}, {Pauwels}, {Petit}, {Recio-Blanco}, {Richards},
  {Rimoldini}, {Robin}, {Sarro}, {Siopis}, {Smith}, {Sozzetti}, {S{\"u}veges},
  {Torra}, {van Reeven}, {Abbas}, {Abreu Aramburu}, {Accart}, {Aerts},
  {Altavilla}, {{\'A}lvarez}, {Alvarez}, {Alves}, {Anderson}, {Andrei},
  {Anglada Varela}, {Antiche}, {Antoja}, {Arcay}, {Astraatmadja}, {Bach},
  {Baker}, {Balaguer-N{\'u}{\~n}ez}, {Balm}, {Barache}, {Barata}, {Barbato},
  {Barblan}, {Barklem}, {Barrado}, {Barros}, {Barstow}, {Bartholom{\'e}
  Mu{\~n}oz}, {Bassilana}, {Becciani}, {Bellazzini}, {Berihuete}, {Bertone},
  {Bianchi}, {Bienaym{\'e}}, {Blanco-Cuaresma}, {Boch}, {Boeche}, {Bombrun},
  {Borrachero}, {Bossini}, {Bouquillon}, {Bourda}, {Bragaglia}, {Bramante},
  {Breddels}, {Bressan}, {Brouillet}, {Br{\"u}semeister}, {Brugaletta},
  {Bucciarelli}, {Burlacu}, {Busonero}, {Butkevich}, {Buzzi}, {Caffau},
  {Cancelliere}, {Cannizzaro}, {Cantat-Gaudin}, {Carballo}, {Carlucci},
  {Carrasco}, {Casamiquela}, {Castellani}, {Castro-Ginard}, {Charlot},
  {Chemin}, {Chiavassa}, {Cocozza}, {Costigan}, {Cowell}, {Crifo}, {Crosta},
  {Crowley}, {Cuypers}, {Dafonte}, {Damerdji}, {Dapergolas}, {David}, {David},
  {de Laverny}, {De Luise}, {De March}, {de Martino}, {de Souza}, {de Torres},
  {Debosscher}, {del Pozo}, {Delbo}, {Delgado}, {Delgado}, {Di Matteo},
  {Diakite}, {Diener}, {Distefano}, {Dolding}, {Drazinos}, {Dur{\'a}n},
  {Edvardsson}, {Enke}, {Eriksson}, {Esquej}, {Eynard Bontemps}, {Fabre},
  {Fabrizio}, {Faigler}, {Falc{\~a}o}, {Farr{\`a}s Casas}, {Federici},
  {Fedorets}, {Fernique}, {Figueras}, {Filippi}, {Findeisen}, {Fonti},
  {Fraile}, {Fraser}, {Fr{\'e}zouls}, {Gai}, {Galleti}, {Garabato},
  {Garc{\'\i}a-Sedano}, {Garofalo}, {Garralda}, {Gavel}, {Gavras}, {Gerssen},
  {Geyer}, {Giacobbe}, {Gilmore}, {Girona}, {Giuffrida}, {Glass}, {Gomes},
  {Granvik}, {Gueguen}, {Guerrier}, {Guiraud}, {Guti{\'e}rrez-S{\'a}nchez},
  {Haigron}, {Hatzidimitriou}, {Hauser}, {Haywood}, {Heiter}, {Helmi}, {Heu},
  {Hilger}, {Hobbs}, {Hofmann}, {Holland}, {Huckle}, {Hypki}, {Icardi},
  {Jan{\ss}en}, {Jevardat de Fombelle}, {Jonker}, {Juh{\'a}sz}, {Julbe},
  {Karampelas}, {Kewley}, {Klar}, {Kochoska}, {Kohley}, {Kolenberg},
  {Kontizas}, {Kontizas}, {Koposov}, {Kordopatis}, {Kostrzewa-Rutkowska},
  {Koubsky}, {Lambert}, {Lanza}, {Lasne}, {Lavigne}, {Le Fustec}, {Le
  Poncin-Lafitte}, {Lebreton}, {Leccia}, {Leclerc}, {Lecoeur-Taibi},
  {Lenhardt}, {Leroux}, {Liao}, {Licata}, {Lindstr{\o}m}, {Lister}, {Livanou},
  {Lobel}, {L{\'o}pez}, {Managau}, {Mann}, {Mantelet}, {Marchal}, {Marchant},
  {Marconi}, {Marinoni}, {Marschalk{\'o}}, {Marshall}, {Martino}, {Marton},
  {Mary}, {Massari}, {Matijevi{\v{c}}}, {Mazeh}, {McMillan}, {Messina},
  {Michalik}, {Millar}, {Molina}, {Molinaro}, {Moln{\'a}r}, {Montegriffo},
  {Mor}, {Morbidelli}, {Morel}, {Morris}, {Mulone}, {Muraveva}, {Musella},
  {Nelemans}, {Nicastro}, {Noval}, {O'Mullane}, {Ord{\'e}novic},
  {Ord{\'o}{\~n}ez-Blanco}, {Osborne}, {Pagani}, {Pagano}, {Pailler},
  {Palacin}, {Palaversa}, {Panahi}, {Pawlak}, {Piersimoni}, {Pineau}, {Plachy},
  {Plum}, {Poggio}, {Poujoulet}, {Pr{\v{s}}a}, {Pulone}, {Racero}, {Ragaini},
  {Rambaux}, {Ramos-Lerate}, {Regibo}, {Reyl{\'e}}, {Riclet}, {Ripepi}, {Riva},
  {Rivard}, {Rixon}, {Roegiers}, {Roelens}, {Romero-G{\'o}mez}, {Rowell},
  {Royer}, {Ruiz-Dern}, {Sadowski}, {Sagrist{\`a} Sell{\'e}s}, {Sahlmann},
  {Salgado}, {Salguero}, {Sanna}, {Santana-Ros}, {Sarasso}, {Savietto},
  {Schultheis}, {Sciacca}, {Segol}, {Segovia}, {S{\'e}gransan}, {Shih},
  {Siltala}, {Silva}, {Smart}, {Smith}, {Solano}, {Solitro}, {Sordo}, {Soria
  Nieto}, {Souchay}, {Spagna}, {Spoto}, {Stampa}, {Steele},
  {Steidelm{\"u}ller}, {Stephenson}, {Stoev}, {Suess}, {Surdej}, {Szabados},
  {Szegedi-Elek}, {Tapiador}, {Taris}, {Tauran}, {Taylor}, {Teixeira},
  {Terrett}, {Teyssandier}, {Thuillot}, {Titarenko}, {Torra Clotet}, {Turon},
  {Ulla}, {Utrilla}, {Uzzi}, {Vaillant}, {Valentini}, {Valette}, {van Elteren},
  {Van Hemelryck}, {van Leeuwen}, {Vaschetto}, {Vecchiato}, {Veljanoski},
  {Viala}, {Vicente}, {Vogt}, {von Essen}, {Voss}, {Votruba}, {Voutsinas},
  {Walmsley}, {Weiler}, {Wertz}, {Wevers}, {Wyrzykowski}, {Yoldas},
  {{\v{Z}}erjal}, {Ziaeepour}, {Zorec}, {Zschocke}, {Zucker}, {Zurbach}, \&
  {Zwitter}}]{GaiaDR2}
{Gaia Collaboration}, {Brown}, A.~G.~A., {Vallenari}, A., {et~al.} 2018, \aap,
  616, A1, \dodoi{10.1051/0004-6361/201833051}

\bibitem[{{Gandhi} {et~al.}(2019){Gandhi}, {Madhusudhan}, {Hawker}, \&
  {Piette}}]{Gandhi19}
{Gandhi}, S., {Madhusudhan}, N., {Hawker}, G., \& {Piette}, A. 2019, \aj, 158,
  228, \dodoi{10.3847/1538-3881/ab4efc}

\bibitem[{{Gandhi} {et~al.}(2020{\natexlab{a}}){Gandhi}, {Madhusudhan}, \&
  {Mandell}}]{Gandhi2020}
{Gandhi}, S., {Madhusudhan}, N., \& {Mandell}, A. 2020{\natexlab{a}}, \aj, 159,
  232, \dodoi{10.3847/1538-3881/ab845e}

\bibitem[{{Gandhi} {et~al.}(2020{\natexlab{b}}){Gandhi}, {Brogi}, {Yurchenko},
  {Tennyson}, {Coles}, {Webb}, {Birkby}, {Guilluy}, {Hawker}, {Madhusudhan},
  {Bonomo}, \& {Sozzetti}}]{Gandhi2020xsec}
{Gandhi}, S., {Brogi}, M., {Yurchenko}, S.~N., {et~al.} 2020{\natexlab{b}},
  \mnras, 495, 224, \dodoi{10.1093/mnras/staa981}

\bibitem[{{Gharib-Nezhad} {et~al.}(2021){Gharib-Nezhad}, {Iyer}, {Line},
  {Freedman}, {Marley}, \& {Batalha}}]{GharibNezhad21}
{Gharib-Nezhad}, E., {Iyer}, A.~R., {Line}, M.~R., {et~al.} 2021, arXiv
  e-prints, arXiv:2104.00264.
\newblock \doarXiv{2104.00264}

\bibitem[{{Giacobbe} {et~al.}(2021){Giacobbe}, {Brogi}, {Gandhi}, {Cubillos},
  {Bonomo}, {Sozzetti}, {Fossati}, {Guilluy}, {Carleo}, {Rainer},
  {Harutyunyan}, {Borsa}, {Pino}, {Nascimbeni}, {Benatti}, {Biazzo},
  {Bignamini}, {Chubb}, {Claudi}, {Cosentino}, {Covino}, {Damasso}, {Desidera},
  {Fiorenzano}, {Ghedina}, {Lanza}, {Leto}, {Maggio}, {Malavolta}, {Maldonado},
  {Micela}, {Molinari}, {Pagano}, {Pedani}, {Piotto}, {Poretti}, {Scandariato},
  {Yurchenko}, {Fantinel}, {Galli}, {Lodi}, {Sanna}, \& {Tozzi}}]{Giacobbe2021}
{Giacobbe}, P., {Brogi}, M., {Gandhi}, S., {et~al.} 2021, \nat, 592, 205,
  \dodoi{10.1038/s41586-021-03381-x}

\bibitem[{Gordon {et~al.}(2022)Gordon, Rothman, Hargreaves, Hashemi, Karlovets,
  Skinner, Conway, Hill, Kochanov, Tan, {et~al.}}]{gordon2022hitran2020}
Gordon, I., Rothman, L., Hargreaves, R., {et~al.} 2022, Journal of quantitative
  spectroscopy and radiative transfer, 277, 107949

\bibitem[{{Grevesse} {et~al.}(2007){Grevesse}, {Asplund}, \&
  {Sauval}}]{Grevesse2007}
{Grevesse}, N., {Asplund}, M., \& {Sauval}, A.~J. 2007, \ssr, 130, 105,
  \dodoi{10.1007/s11214-007-9173-7}

\bibitem[{{Grimm} {et~al.}(2021){Grimm}, {Malik}, {Kitzmann},
  {Guzm{\'a}n-Mesa}, {Hoeijmakers}, {Fisher}, {Mendon{\c{c}}a}, {Yurchenko},
  {Tennyson}, {Alesina}, {Buchschacher}, {Burnier}, {Segransan}, {Kurucz}, \&
  {Heng}}]{Grimm2021}
{Grimm}, S.~L., {Malik}, M., {Kitzmann}, D., {et~al.} 2021, \apjs, 253, 30,
  \dodoi{10.3847/1538-4365/abd773}

\bibitem[{{Guillot}(2010)}]{Guillot2010}
{Guillot}, T. 2010, \aap, 520, A27+, \dodoi{10.1051/0004-6361/200913396}

\bibitem[{{Guillot} {et~al.}(2022){Guillot}, {Fletcher}, {Helled}, {Ikoma},
  {Line}, \& {Parmentier}}]{Guillot2022}
{Guillot}, T., {Fletcher}, L.~N., {Helled}, R., {et~al.} 2022, arXiv e-prints,
  arXiv:2205.04100.
\newblock \doarXiv{2205.04100}

\bibitem[{Harris {et~al.}(2020)Harris, Millman, van~der Walt, Gommers,
  Virtanen, Cournapeau, Wieser, Taylor, Berg, Smith, Kern, Picus, Hoyer, van
  Kerkwijk, Brett, Haldane, del R{\'{i}}o, Wiebe, Peterson,
  G{\'{e}}rard-Marchant, Sheppard, Reddy, Weckesser, Abbasi, Gohlke, \&
  Oliphant}]{numpy:2020}
Harris, C.~R., Millman, K.~J., van~der Walt, S.~J., {et~al.} 2020, Nature, 585,
  357, \dodoi{10.1038/s41586-020-2649-2}

\bibitem[{{Hellier} {et~al.}(2009){Hellier}, {Anderson}, {Collier Cameron},
  {Gillon}, {Hebb}, {Maxted}, {Queloz}, {Smalley}, {Triaud}, {West}, {Wilson},
  {Bentley}, {Enoch}, {Horne}, {Irwin}, {Lister}, {Mayor}, {Parley}, {Pepe},
  {Pollacco}, {Segransan}, {Udry}, \& {Wheatley}}]{Hellier2009}
{Hellier}, C., {Anderson}, D.~R., {Collier Cameron}, A., {et~al.} 2009, \nat,
  460, 1098, \dodoi{10.1038/nature08245}

\bibitem[{{Hubeny} {et~al.}(2003){Hubeny}, {Burrows}, \&
  {Sudarsky}}]{Hubeny2003}
{Hubeny}, I., {Burrows}, A., \& {Sudarsky}, D. 2003, \apj, 594, 1011,
  \dodoi{10.1086/377080}

\bibitem[{Hunter(2007)}]{matplotlib:2007}
Hunter, J.~D. 2007, Computing in Science \& Engineering, 9, 90,
  \dodoi{10.1109/MCSE.2007.55}

\bibitem[{{Iro} {et~al.}(2005){Iro}, {B{\'e}zard}, \& {Guillot}}]{Iro2005}
{Iro}, N., {B{\'e}zard}, B., \& {Guillot}, T. 2005, \aap, 436, 719,
  \dodoi{10.1051/0004-6361:20048344}

\bibitem[{{John}(1988)}]{John1988}
{John}, T.~L. 1988, \aap, 193, 189

\bibitem[{Kanodia \& Wright(2018)}]{barycorrpy}
Kanodia, S., \& Wright, J. 2018, Research Notes of the {AAS}, 2, 4,
  \dodoi{10.3847/2515-5172/aaa4b7}

\bibitem[{{Kasper} {et~al.}(2021){Kasper}, {Bean}, {Line}, {Seifahrt},
  {St{\"u}rmer}, {Pino}, {D{\'e}sert}, \& {Brogi}}]{kasper2021}
{Kasper}, D., {Bean}, J.~L., {Line}, M.~R., {et~al.} 2021, \apjl, 921, L18,
  \dodoi{10.3847/2041-8213/ac30e1}

\bibitem[{{Kasper} {et~al.}(2022){Kasper}, {Bean}, {Line}, {Seifahrt},
  {Lothringer}, {Pino}, {Fu}, {Pelletier}, {St{\"u}rmer}, {Bj{\"o}rn Benneke},
  {Brogi}, \& {D{\'e}sert}}]{Kasper2022}
---. 2022, arXiv e-prints, arXiv:2208.04759.
\newblock \doarXiv{2208.04759}

\bibitem[{{Kempton} {et~al.}(2017){Kempton}, {Bean}, \&
  {Parmentier}}]{Kempton2017}
{Kempton}, E.~M.-R., {Bean}, J.~L., \& {Parmentier}, V. 2017, \apjl, 845, L20,
  \dodoi{10.3847/2041-8213/aa84ac}

\bibitem[{{Kesseli} {et~al.}(2022){Kesseli}, {Snellen}, {Casasayas-Barris},
  {Molli{\`e}re}, \& {S{\'a}nchez-L{\'o}pez}}]{Kesseli2022}
{Kesseli}, A.~Y., {Snellen}, I.~A.~G., {Casasayas-Barris}, N., {Molli{\`e}re},
  P., \& {S{\'a}nchez-L{\'o}pez}, A. 2022, \aj, 163, 107,
  \dodoi{10.3847/1538-3881/ac4336}

\bibitem[{{Kolecki} \& {Wang}(2022)}]{KoleckiWang22}
{Kolecki}, J.~R., \& {Wang}, J. 2022, \aj, 164, 87,
  \dodoi{10.3847/1538-3881/ac7de3}

\bibitem[{{Kreidberg} {et~al.}(2015){Kreidberg}, {Line}, {Bean}, {Stevenson},
  {D{\'e}sert}, {Madhusudhan}, {Fortney}, {Barstow}, {Henry}, {Williamson}, \&
  {Showman}}]{kreidberg2015}
{Kreidberg}, L., {Line}, M.~R., {Bean}, J.~L., {et~al.} 2015, \apj, 814, 66,
  \dodoi{10.1088/0004-637X/814/1/66}

\bibitem[{{Kurucz} \& {Bell}(1995)}]{kurucz1995}
{Kurucz}, R.~L., \& {Bell}, B. 1995, {Atomic line list} (CD-ROM No. 23,
  Cambridge, MA: Smithsonian Astrophysical Observatory)

\bibitem[{{Lee} \& {Gullikson}(2016)}]{LeePLP2016}
{Lee}, J.-J., \& {Gullikson}, K. 2016, {Plp: V2.1 Alpha 3}, v2.1-alpha.3,
  Zenodo, \dodoi{10.5281/zenodo.56067}

\bibitem[{{Line} {et~al.}(2013){Line}, {Knutson}, {Deming}, {Wilkins}, \&
  {Desert}}]{Line2013a}
{Line}, M.~R., {Knutson}, H., {Deming}, D., {Wilkins}, A., \& {Desert}, J.-M.
  2013, \apj, 778, 183, \dodoi{10.1088/0004-637X/778/2/183}

\bibitem[{{Line} {et~al.}(2015){Line}, {Teske}, {Burningham}, {Fortney}, \&
  {Marley}}]{Line2015}
{Line}, M.~R., {Teske}, J., {Burningham}, B., {Fortney}, J.~J., \& {Marley},
  M.~S. 2015, \apj, 807, 183, \dodoi{10.1088/0004-637X/807/2/183}

\bibitem[{{Line} {et~al.}(2021){Line}, {Brogi}, {Bean}, {Gandhi}, {Zalesky},
  {Parmentier}, {Smith}, {Mace}, {Mansfield}, {Kempton}, {Fortney}, {Shkolnik},
  {Patience}, {Rauscher}, {D{\'e}sert}, \& {Wardenier}}]{line2021}
{Line}, M.~R., {Brogi}, M., {Bean}, J.~L., {et~al.} 2021, \nat, 598, 580,
  \dodoi{10.1038/s41586-021-03912-6}

\bibitem[{Lodders {et~al.}(2009)Lodders, Palme, \& Gail}]{Lodders2009}
Lodders, K., Palme, H., \& Gail, H.-P. 2009, 712--770,
  \dodoi{10.1007/978-3-540-88055-4_34}

\bibitem[{{Lothringer} {et~al.}(2018){Lothringer}, {Barman}, \&
  {Koskinen}}]{lothringer18}
{Lothringer}, J.~D., {Barman}, T., \& {Koskinen}, T. 2018, \apj, 866, 27,
  \dodoi{10.3847/1538-4357/aadd9e}

\bibitem[{{Mace} {et~al.}(2018){Mace}, {Sokal}, {Lee}, {Oh}, {Park}, {Lee},
  {Good}, {MacQueen}, {Oh}, {Kaplan}, {Kidder}, {Chun}, {Yuk}, {Jeong}, {Pak},
  {Kim}, {Nah}, {Lee}, {Yu}, {Hwang}, {Park}, {Kim}, {Chinn}, {Peck}, {Diaz},
  {Rutten}, {Prato}, {Jacoby}, {Cornelius}, {Hardesty}, {DeGroff}, {Dunham},
  {Levine}, {Nofi}, {Lopez-Valdivia}, {Weinberger}, \& {Jaffe}}]{Mace2018}
{Mace}, G., {Sokal}, K., {Lee}, J.-J., {et~al.} 2018, in Society of
  Photo-Optical Instrumentation Engineers (SPIE) Conference Series, Vol. 10702,
  Ground-based and Airborne Instrumentation for Astronomy VII, ed. C.~J.
  {Evans}, L.~{Simard}, \& H.~{Takami}, 107020Q, \dodoi{10.1117/12.2312345}

\bibitem[{{Madhusudhan}(2018)}]{Madhu2018}
{Madhusudhan}, N. 2018, in Handbook of Exoplanets, ed. H.~J. {Deeg} \& J.~A.
  {Belmonte}, 104, \dodoi{10.1007/978-3-319-55333-7\_104}

\bibitem[{{Madhusudhan}(2019)}]{Madhusudhan19}
---. 2019, \araa, 57, 617, \dodoi{10.1146/annurev-astro-081817-051846}

\bibitem[{{Madhusudhan} {et~al.}(2014){Madhusudhan}, {Crouzet}, {McCullough},
  {Deming}, \& {Hedges}}]{Madhusudhan2014a}
{Madhusudhan}, N., {Crouzet}, N., {McCullough}, P.~R., {Deming}, D., \&
  {Hedges}, C. 2014, \apjl, 791, L9, \dodoi{10.1088/2041-8205/791/1/L9}

\bibitem[{{Madhusudhan} {et~al.}(2011){Madhusudhan}, {Mousis}, {Johnson}, \&
  {Lunine}}]{Madhusudhan2011}
{Madhusudhan}, N., {Mousis}, O., {Johnson}, T.~V., \& {Lunine}, J.~I. 2011,
  \apj, 743, 191, \dodoi{10.1088/0004-637X/743/2/191}

\bibitem[{{Madhusudhan} \& {Seager}(2009)}]{Madhusudhan2009}
{Madhusudhan}, N., \& {Seager}, S. 2009, \apj, 707, 24,
  \dodoi{10.1088/0004-637X/707/1/24}

\bibitem[{{Mansfield} {et~al.}(2021){Mansfield}, {Line}, {Bean}, {Fortney},
  {Parmentier}, {Wiser}, {Kempton}, {Gharib-Nezhad}, {Sing},
  {L{\'o}pez-Morales}, {Baxter}, {D{\'e}sert}, {Swain}, \&
  {Roudier}}]{Mansfield2021}
{Mansfield}, M., {Line}, M.~R., {Bean}, J.~L., {et~al.} 2021, Nature Astronomy,
  5, 1224, \dodoi{10.1038/s41550-021-01455-4}

\bibitem[{{Maxted} {et~al.}(2013){Maxted}, {Anderson}, {Doyle}, {Gillon},
  {Harrington}, {Iro}, {Jehin}, {Lafreni{\`e}re}, {Smalley}, \&
  {Southworth}}]{Maxted2013}
{Maxted}, P.~F.~L., {Anderson}, D.~R., {Doyle}, A.~P., {et~al.} 2013, \mnras,
  428, 2645, \dodoi{10.1093/mnras/sts231}

\bibitem[{{Nymeyer} {et~al.}(2011){Nymeyer}, {Harrington}, {Hardy},
  {Stevenson}, {Campo}, {Madhusudhan}, {Collier-Cameron}, {Loredo}, {Blecic},
  {Bowman}, {Britt}, {Cubillos}, {Hellier}, {Gillon}, {Maxted}, {Hebb},
  {Wheatley}, {Pollacco}, \& {Anderson}}]{Nymeyer2011}
{Nymeyer}, S., {Harrington}, J., {Hardy}, R.~A., {et~al.} 2011, \apj, 742, 35,
  \dodoi{10.1088/0004-637X/742/1/35}

\bibitem[{{Park} {et~al.}(2014){Park}, {Jaffe}, {Yuk}, {Chun}, {Pak}, {Kim},
  {Pavel}, {Lee}, {Oh}, {Jeong}, {Sim}, {Lee}, {Nguyen Le}, {Strubhar},
  {Gully-Santiago}, {Oh}, {Cha}, {Moon}, {Park}, {Brooks}, {Ko}, {Han}, {Nah},
  {Hill}, {Lee}, {Barnes}, {Yu}, {Kaplan}, {Mace}, {Kim}, {Lee}, {Hwang}, \&
  {Park}}]{Park2014}
{Park}, C., {Jaffe}, D.~T., {Yuk}, I.-S., {et~al.} 2014, in Society of
  Photo-Optical Instrumentation Engineers (SPIE) Conference Series, Vol. 9147,
  Ground-based and Airborne Instrumentation for Astronomy V, ed. S.~K.
  {Ramsay}, I.~S. {McLean}, \& H.~{Takami}, 91471D, \dodoi{10.1117/12.2056431}

\bibitem[{{Parmentier} {et~al.}(2021){Parmentier}, {Showman}, \&
  {Fortney}}]{Parmentier2021}
{Parmentier}, V., {Showman}, A.~P., \& {Fortney}, J.~J. 2021, \mnras, 501, 78,
  \dodoi{10.1093/mnras/staa3418}

\bibitem[{{Parmentier} {et~al.}(2018{\natexlab{a}}){Parmentier}, {Line},
  {Bean}, {Mansfield}, {Kreidberg}, {Lupu}, {Visscher}, {D{\'e}sert},
  {Fortney}, {Deleuil}, {Arcangeli}, {Showman}, \& {Marley}}]{parmentier18}
{Parmentier}, V., {Line}, M.~R., {Bean}, J.~L., {et~al.} 2018{\natexlab{a}},
  \aap, 617, A110, \dodoi{10.1051/0004-6361/201833059}

\bibitem[{{Parmentier} {et~al.}(2018{\natexlab{b}}){Parmentier}, {Line},
  {Bean}, {Mansfield}, {Kreidberg}, {Lupu}, {Visscher}, {D{\'e}sert},
  {Fortney}, \& {Deleuil}}]{Parmentier2018}
---. 2018{\natexlab{b}}, \aap, 617, A110, \dodoi{10.1051/0004-6361/201833059}

\bibitem[{{Perez-Becker} \& {Showman}(2013)}]{Perez-Becker2013a}
{Perez-Becker}, D., \& {Showman}, A.~P. 2013, \apj, 776, 134,
  \dodoi{10.1088/0004-637X/776/2/134}

\bibitem[{{Piette} \& {Madhusudhan}(2020)}]{Piette2020}
{Piette}, A. A.~A., \& {Madhusudhan}, N. 2020, \mnras, 497, 5136,
  \dodoi{10.1093/mnras/staa2289}

\bibitem[{{Piskorz} {et~al.}(2018){Piskorz}, {Buzard}, {Line}, {Knutson},
  {Benneke}, {Crockett}, {Lockwood}, {Blake}, {Barman}, {Bender}, {Deming}, \&
  {Johnson}}]{pis18}
{Piskorz}, D., {Buzard}, C., {Line}, M.~R., {et~al.} 2018, \aj, 156, 133,
  \dodoi{10.3847/1538-3881/aad781}

\bibitem[{{Polanski} {et~al.}(2022){Polanski}, {Crossfield}, {Howard},
  {Isaacson}, \& {Rice}}]{Polanski2022}
{Polanski}, A.~S., {Crossfield}, I. J.~M., {Howard}, A.~W., {Isaacson}, H., \&
  {Rice}, M. 2022, Research Notes of the American Astronomical Society, 6, 155,
  \dodoi{10.3847/2515-5172/ac8676}

\bibitem[{{Polyansky} {et~al.}(2018){Polyansky}, {Kyuberis}, {Zobov},
  {Tennyson}, {Yurchenko}, \& {Lodi}}]{Polyansky2018}
{Polyansky}, O.~L., {Kyuberis}, A.~A., {Zobov}, N.~F., {et~al.} 2018, \mnras,
  480, 2597, \dodoi{10.1093/mnras/sty1877}

\bibitem[{{Prinoth} {et~al.}(2022){Prinoth}, {Hoeijmakers}, {Kitzmann},
  {Sandvik}, {Seidel}, {Lendl}, {Borsato}, {Thorsbro}, {Anderson}, {Barrado},
  {Kravchenko}, {Allart}, {Bourrier}, {Cegla}, {Ehrenreich}, {Fisher}, {Lovis},
  {Guzm{\'a}n-Mesa}, {Grimm}, {Hooton}, {Morris}, {Oreshenko}, {Pino}, \&
  {Heng}}]{prinoth22}
{Prinoth}, B., {Hoeijmakers}, H.~J., {Kitzmann}, D., {et~al.} 2022, Nature
  Astronomy, 6, 449, \dodoi{10.1038/s41550-021-01581-z}

\bibitem[{{Rothman} {et~al.}(2010){Rothman}, {Gordon}, {Barber}, {Dothe},
  {Gamache}, {Goldman}, {Perevalov}, {Tashkun}, \& {Tennyson}}]{hitemp2010}
{Rothman}, L.~S., {Gordon}, I.~E., {Barber}, R.~J., {et~al.} 2010, \jqsrt, 111,
  2139, \dodoi{10.1016/j.jqsrt.2010.05.001}

\bibitem[{{S{\'a}nchez-L{\'o}pez} {et~al.}(2022){S{\'a}nchez-L{\'o}pez},
  {Landman}, {Molli{\`e}re}, {Casasayas-Barris}, {Kesseli}, \&
  {Snellen}}]{sanchez-lopez2022}
{S{\'a}nchez-L{\'o}pez}, A., {Landman}, R., {Molli{\`e}re}, P., {et~al.} 2022,
  \aap, 661, A78, \dodoi{10.1051/0004-6361/202142591}

\bibitem[{{Seager}(2010)}]{SeagerBook2010}
{Seager}, S. 2010, {Exoplanet Atmospheres: Physical Processes}

\bibitem[{{Sheppard} {et~al.}(2017){Sheppard}, {Mandell}, {Tamburo}, {Gandhi},
  {Pinhas}, {Madhusudhan}, \& {Deming}}]{Sheppard2017}
{Sheppard}, K.~B., {Mandell}, A.~M., {Tamburo}, P., {et~al.} 2017, \apjl, 850,
  L32, \dodoi{10.3847/2041-8213/aa9ae9}

\bibitem[{{Shporer} {et~al.}(2019){Shporer}, {Wong}, {Huang}, {Line},
  {Stassun}, {Fetherolf}, {Kane}, {Bouma}, {Daylan}, {G{\"u}enther}, {Ricker},
  {Latham}, {Vanderspek}, {Seager}, {Winn}, {Jenkins}, {Glidden},
  {Berta-Thompson}, {Ting}, {Li}, \& {Haworth}}]{Shporer2019}
{Shporer}, A., {Wong}, I., {Huang}, C.~X., {et~al.} 2019, \aj, 157, 178,
  \dodoi{10.3847/1538-3881/ab0f96}

\bibitem[{{Sing} {et~al.}(2016){Sing}, {Fortney}, {Nikolov}, {Wakeford},
  {Kataria}, {Evans}, {Aigrain}, {Ballester}, {Burrows}, {Deming},
  {D{\'e}sert}, {Gibson}, {Henry}, {Huitson}, {Knutson}, {Etangs}, {Pont},
  {Showman}, {Vidal-Madjar}, {Williamson}, \& {Wilson}}]{Sing2016}
{Sing}, D.~K., {Fortney}, J.~J., {Nikolov}, N., {et~al.} 2016, \nat, 529, 59,
  \dodoi{10.1038/nature16068}

\bibitem[{{Snellen} {et~al.}(2010){Snellen}, {de Kok}, {de Mooij}, \&
  {Albrecht}}]{Snellen2010}
{Snellen}, I.~A.~G., {de Kok}, R.~J., {de Mooij}, E.~J.~W., \& {Albrecht}, S.
  2010, \nat, 465, 1049, \dodoi{10.1038/nature09111}

\bibitem[{{Stangret} {et~al.}(2020){Stangret}, {Casasayas-Barris}, {Pall{\'e}},
  {Yan}, {S{\'a}nchez-L{\'o}pez}, \& {L{\'o}pez-Puertas}}]{Stangret20}
{Stangret}, M., {Casasayas-Barris}, N., {Pall{\'e}}, E., {et~al.} 2020, \aap,
  638, A26, \dodoi{10.1051/0004-6361/202037541}

\bibitem[{{Tabernero} {et~al.}(2021){Tabernero}, {Zapatero Osorio}, {Allart},
  {Borsa}, {Casasayas-Barris}, {Demangeon}, {Ehrenreich}, {Lillo-Box}, {Lovis},
  {Pall{\'e}}, {Sousa}, {Rebolo}, {Santos}, {Pepe}, {Cristiani}, {Adibekyan},
  {Allende Prieto}, {Alibert}, {Barros}, {Bouchy}, {Bourrier}, {D'Odorico},
  {Dumusque}, {Faria}, {Figueira}, {G{\'e}nova Santos}, {Gonz{\'a}lez
  Hern{\'a}ndez}, {Hojjatpanah}, {Lo Curto}, {Lavie}, {Martins}, {Martins},
  {Mehner}, {Micela}, {Molaro}, {Nunes}, {Poretti}, {Seidel}, {Sozzetti},
  {Su{\'a}rez Mascare{\~n}o}, {Udry}, {Aliverti}, {Affolter}, {Alves}, {Amate},
  {Avila}, {Bandy}, {Benz}, {Bianco}, {Broeg}, {Cabral}, {Conconi}, {Coelho},
  {Cumani}, {Deiries}, {Dekker}, {Delabre}, {Fragoso}, {Genoni}, {Genolet},
  {Hughes}, {Knudstrup}, {Kerber}, {Landoni}, {Lizon}, {Maire}, {Manescau}, {Di
  Marcantonio}, {M{\'e}gevand}, {Monteiro}, {Monteiro}, {Moschetti}, {Mueller},
  {Modigliani}, {Oggioni}, {Oliveira}, {Pariani}, {Pasquini}, {Rasilla},
  {Redaelli}, {Riva}, {Santana-Tschudi}, {Santin}, {Santos}, {Segovia},
  {Sosnowska}, {Span{\`o}}, {Tenegi}, {Iwert}, {Zanutta}, \&
  {Zerbi}}]{Tabernero2021}
{Tabernero}, H.~M., {Zapatero Osorio}, M.~R., {Allart}, R., {et~al.} 2021,
  \aap, 646, A158, \dodoi{10.1051/0004-6361/202039511}

\bibitem[{{Tannock} {et~al.}(2022){Tannock}, {Metchev}, {Hood}, {Mace},
  {Fortney}, {Morley}, {Jaffe}, \& {Lupu}}]{Tannock2022}
{Tannock}, M.~E., {Metchev}, S., {Hood}, C.~E., {et~al.} 2022, \mnras, 514,
  3160, \dodoi{10.1093/mnras/stac1412}

\bibitem[{{Tennyson} {et~al.}(2020){Tennyson}, {Yurchenko}, {Al-Refaie},
  {Clark}, {Chubb}, {Conway}, {Dewan}, {Gorman}, {Hill}, {Lynas-Gray},
  {Mellor}, {McKemmish}, {Owens}, {Polyansky}, {Semenov}, {Somogyi}, {Tinetti},
  {Upadhyay}, {Waldmann}, {Wang}, {Wright}, \& {Yurchenko}}]{Tennyson2020}
{Tennyson}, J., {Yurchenko}, S.~N., {Al-Refaie}, A.~F., {et~al.} 2020, arXiv
  e-prints, arXiv:2007.13022.
\newblock \doarXiv{2007.13022}

\bibitem[{{Thorngren} {et~al.}(2016){Thorngren}, {Fortney}, {Murray-Clay}, \&
  {Lopez}}]{Thorngren2016}
{Thorngren}, D.~P., {Fortney}, J.~J., {Murray-Clay}, R.~A., \& {Lopez}, E.~D.
  2016, \apj, 831, 64, \dodoi{10.3847/0004-637X/831/1/64}

\bibitem[{{Trotta}(2008)}]{trotta2008bayes}
{Trotta}, R. 2008, Contemporary Physics, 49, 71,
  \dodoi{10.1080/00107510802066753}

\bibitem[{Van~Rossum \& Drake(2009)}]{python3:2009}
Van~Rossum, G., \& Drake, F.~L. 2009, Python 3 Reference Manual (Scotts Valley,
  CA: CreateSpace)

\bibitem[{{van Sluijs} {et~al.}(2022){van Sluijs}, {Birkby}, {Lothringer},
  {Lee}, {Crossfield}, {Parmentier}, {Brogi}, {Kulesa}, {McCarthy}, {Powell},
  \& {Charbonneau}}]{vansluijs22}
{van Sluijs}, L., {Birkby}, J.~L., {Lothringer}, J., {et~al.} 2022, arXiv
  e-prints, arXiv:2203.13234.
\newblock \doarXiv{2203.13234}

\bibitem[{Virtanen {et~al.}(2020)Virtanen, Gommers, Oliphant, Haberland, Reddy,
  Cournapeau, Burovski, Peterson, Weckesser, Bright, {van der Walt}, Brett,
  Wilson, Millman, Mayorov, Nelson, Jones, Kern, Larson, Carey, Polat, Feng,
  Moore, {VanderPlas}, Laxalde, Perktold, Cimrman, Henriksen, Quintero, Harris,
  Archibald, Ribeiro, Pedregosa, {van Mulbregt}, \& {SciPy 1.0
  Contributors}}]{2020SciPy-NMeth}
Virtanen, P., Gommers, R., Oliphant, T.~E., {et~al.} 2020, Nature Methods, 17,
  261, \dodoi{10.1038/s41592-019-0686-2}

\bibitem[{{Welbanks} {et~al.}(2019){Welbanks}, {Madhusudhan}, {Allard},
  {Hubeny}, {Spiegelman}, \& {Leininger}}]{Welbanks2019}
{Welbanks}, L., {Madhusudhan}, N., {Allard}, N.~F., {et~al.} 2019, \apjl, 887,
  L20, \dodoi{10.3847/2041-8213/ab5a89}

\bibitem[{{Wong} {et~al.}(2020){Wong}, {Shporer}, {Daylan}, {Benneke},
  {Fetherolf}, {Kane}, {Ricker}, {Vanderspek}, {Latham}, {Winn}, {Jenkins},
  {Boyd}, {Glidden}, {Goeke}, {Sha}, {Ting}, \& {Yahalomi}}]{Wong2020}
{Wong}, I., {Shporer}, A., {Daylan}, T., {et~al.} 2020, \aj, 160, 155,
  \dodoi{10.3847/1538-3881/ababad}

\bibitem[{{Zhang}(2020)}]{ZhangXi2020}
{Zhang}, X. 2020, Research in Astronomy and Astrophysics, 20, 099,
  \dodoi{10.1088/1674-4527/20/7/99}

\end{thebibliography}
\bibliographystyle{aasjournal}

\end{document}